\documentclass[usenatbib,referee]{mn2e}
\usepackage{graphicx}

\title[X-ray Polarization in Magnetar Emission]{Soft X-ray Polarization in Thermal Magnetar Emission}

\author[van Adelsberg \& Perna]
{Matthew van Adelsberg$^1$ \& Rosalba Perna$^2$\\
$^1$Kavli Institute for Theoretical Physics, Kohn Hall,
University of California, Santa Barbara, CA 93106\\
$^2$JILA and Department of Astrophysical
and Planetary Sciences, University of Colorado, Boulder, CO 80309}

\begin{document}

\maketitle

\begin{abstract}
Emission spectra from magnetars in the soft X-ray band likely contain a thermal
component emerging directly from the neutron star surface.  However, the
lack of observed absorption-like features in quiescent spectra makes
it difficult to directly constrain physical properties of the 
atmosphere.  We argue that future X-ray polarization measurements
represent a promising technique for directly constraining the magnetar
magnetic field strength and geometry.  We construct models of the
observed polarization signal from a finite surface hotspot, using the
latest NS atmosphere models for magnetic fields $B = 4\times
10^{13}-5\times 10^{14}$~G.  Our calculations are strongly dependent
on the NS magnetic field strength and geometry, and are more weakly dependent
on the NS equation of state and atmosphere composition.  We discuss
how the complementary dependencies of phase-resolved spectroscopy and
polarimetry might resolve degeneracies that currently hamper the
determination of magnetar physical parameters using thermal models.
\end{abstract}


\section{Introduction}
\label{sect:Introduction}

Thermal models of emission from neutron star (NS) surfaces and
observations of NS cooling can be used to constrain the star equation
of state (EOS), properties of matter at super-nuclear densitites, and
physics at extreme magnetic field strengths
\citep[e.g.,][]{Prakashetal01a,YakovlevPethick04a}.  Thermal radiation
has been detected from several classes of isolated NSs, including
Anomolous X-ray Pulsars (AXPs), Soft Gamma-ray Repeaters (SGRs), X-ray
Dim Isolated Neutron Stars \citep[XDINS; e.g.,][]{Haberl07a}, pulsars
\citep[e.g.,][]{BeckerPavlov02a,MarshallSchulz02a,deLucaetal05a}, and
central compact objects in supernova remnants
\citep[e.g.,][]{Sanwaletal02a,Morietal05a,Kargaltsevetal05a}.  In this
paper, we focus on emission from AXPs and SGRs, which comprise the
population of magnetars, NSs that feature quiescent and bursting
emission powered by the decay of strong interior magnetic fields with
surface strengths $B\sim 10^{14}-10^{15}$~G \citep[e.g.,][and the
references therein]{WoodsThompson06a}.  There is a near consensus in
the literature that the magnetar hypothesis is correct.  Evidence
favoring the magnetar model includes the energetics, flux, and timing
properties of SGR and AXP flares, long magnetar rotation periods and
inferred ages, and the lack of sufficient NS rotational energy to
power the observed quiescent emission \citep[e.g.,][]{Kaspi04a}.  This
phenomenology is consistent with internal NS magnetic fields strengths
of $B\ga 10^{15}$~G.  However, other than the large magnetic dipole
components calculated from timing measurements, there are no direct
empirical constraints on external magnetar-strength magnetic fields. This is in
part due to the lack of features in magnetar thermal spectra, a
characteristic that is difficult to explain, since broad features have
been observed in the spectra of NSs with lower magnetic fields and
temperatures (e.g., XDINS and pulsars).

Moreover, differences between the observed characteristics of magnetars and other classes of high field 
NSs are still poorly understood.  
The recently discovered population of high-field radio pulsars is similar in most respects 
to canonical pulsars (they are powered by rotational kinetic energy and emit persistently in the 
radio band).  However, the high-field radio pulsars have long periods and inferred magnetic field strengths of order  
$B\ga 4\times 10^{13}$~G \citep[e.g.,][]{Camiloetal00a,Gonzalezetal04a,KaspiMcLaughlin05a,Livingstoneetal06a}.
Several authors have speculated that the distinct behaviors of magnetars and high field radio pulsars 
suggest differences in their magnetic field geometries and possible evolutionary 
connections between the two populations \citep[e.g.,][]{Gonzalezetal04a}.
 
One of the keys to understanding the nature of magnetars and their relationship to other NS populations 
is in the production of plausible theoretical models for interpreting observations.  
While significant progress has been made in modeling the spectra of thermal radiation from 
NS atmospheres, the use of such models to explain observational data is still 
in its infancy.  In practice, typical soft X-ray magnetar spectra are fit equally well by blackbody or atmosphere models 
plus a power-law.  The incorporation of vacuum polarization effects into 
NS atmosphere calculations has improved model fits to observed spectra, and has offered a possible explanation for the 
absence of magnetar spectral lines.  In principle, predictions of the beaming pattern of radiation along the surface 
magnetic field can be coupled to phase-resolved spectroscopy to discriminate between NS atmospheres and other emission 
models.  However, the relatively low photon count rates from many magnetar sources, as well as the degeneracies between the 
NS magnetic 
configuration, viewing geometry, and compactness make it difficult to extract physical parameters from observations.

As an alternative, several recent works discuss the evolution of photon polarization states 
in NS magnetospheres, showing that, in the magnetar case, significant linear polarization fractions are expected, 
possibly containing a unique signature of the strong magnetic field  
\citep[][]{HeylShaviv00a,HeylShaviv02a,Heyletal03a,LaiHo03a,vanAdelsbergLai06a,WangLai09a}.
Measurements of significant X-ray polarization, at 2.6~keV and 5.2~keV, were performed for the Crab nebula 
using the 
OSO-8 satellite \citep[][]{Weisskopfetal76a}.  These measurements confirmed an earlier detection by a sounding rocket 
experiment \citep[][]{Novicketal72a}.
However, as of this writing, 
no subsequent 
polarization measurements have been made for any object at energies $E\sim 0.1-10$~keV (relevant for 
thermal magnetar emission). 
Recent advances in instrumentation have stimulated interest in future missions to perform polarimetry in 
the soft X-ray band, leading to several projects which are in active 
development \citep[see][]{Costaetal01a,Kallman04a,Costaetal06a}.

In this paper, we explore the future role of X-ray polarimetry as a complement to spectroscopy in interpreting observational 
spectra.  We will argue that the combination of polarimetry and spectroscopy can 
constrain several critical NS parameters, including the temperature, magnetic field strength and geometry, 
size of emission region, and mass-to-radius ratio of highly magnetized NSs with $B\sim 10^{12}-10^{15}$~G.  
We use the latest magnetar atmosphere models of \citet[][]{vanAdelsbergLai06a}, 
and expand on the work of \citet[][]{HeylShaviv02a}, \citet[][]{Heyletal03a}, \citet[][]{LaiHo03a}, and 
\citet[][]{vanAdelsbergLai06a}, to compute the phase-resolved, 
observed Stokes parameters from magnetars.  We assume that these NSs have   
dipole magnetic field strengths of $B = 4\times 10^{13}$~G -- $5\times 10^{14}$~G and
emit from a region centered around the star polar cap with modest opening angle.  We confirm that the polarization 
signal from a finite region on a magnetar surface retains important information about the strength of the magnetic field, as 
reported in previous works.  We show that this signal has a strong dependence on the magnetic field and viewing geometry, 
and a weaker dependence on the NS EOS, emission region size, and atmosphere composition.  Finally, we argue that polarization measurements can break the degeneracy inherent in inferring physical parameters from spectroscopic measurements alone. 
 
Our paper is organized as follows: in \S\ref{sect:Physics Inputs}, we discuss our assumptions and 
the atmosphere models used 
to calculate the emitted polarization fraction from the NS surface.  
In \S\ref{sect:Emission Model}, we describe our methods for 
calculating the observed polarization signal from an extended NS polar cap, including relativistic effects.  
In \S\ref{sect:Results}, we show the results of our calculations for several representative cases.  Lastly, in 
\S\ref{sect:Discussion}, we analyze our results, and compare and contrast 
the information obtained from polarization studies to that from phase-resolved spectroscopy.

\section{Physics Inputs}
\label{sect:Physics Inputs}

In the canonical pulsar model, the magnetospheres of highly magnetized NSs contain tenuous plasma whose density 
is approximated by the Goldreich-Julian formula,
$n_{\rm GJ} = f B/(c e) \approx 6.9\times 10^{12} f_1 B_{14} \mbox{ cm}^{-3}$, 
where $f_1=f/(\mbox{1 Hz})$ and $B_{14} = B/(10^{14}\mbox{ G})$ are the NS rotation frequency and 
magnetic field strength \citep[e.g.,][]{ShapiroTeukolsky86a}.  
The magnetar model contains a more complicated magnetic field structure compared to the standard dipole case.  
The presence of twisted magnetic fields (e.g., due to crustal motion and magnetic reconfiguration) leads to 
significant magnetospheric 
currents, which can, in principle, maintain electron-positron densities that are orders of 
magnitude larger than $n_{\rm GJ}$ \citep[e.g.,][]{Thompsonetal02a}.  This suggests the 
possibility that resonant cyclotron scattering in the magnetosphere distorts some of the radiation emerging from the 
NS surface, producing a NT component at energies $E\sim 1-10$~keV.  Modeling this emission requires sophisticated 
scattering calculations that depend on the magnetic geometry and magnetosphere structure; significant progress 
has been made in several recent works \citep[][]{LyutikovGavriil06a,FernandezThompson07a,Reaetal08a}.

In this paper, we focus on thermal photons that emerge directly from the NS surface, undistorted by 
scattering processes in the magnetosphere.  In 
addition, we make the simplifying assumption that the NS has a pure dipole magnetic field.  While the actual 
magnetar field structure is likely to be more complicated, it is unlikely to distort the polarization signal, which 
depends on the field structure at distances much greater than the star radius; far from the star, 
the field is dominated by the dipole 
component.  If the emitted polarization is instead determined exclusively by processes at the NS surface or scattering, 
the vacuum polarization signature is destroyed. 

In the work below, we assume that X-ray photons are emitted from a hot region with 
$T_{\rm eff}\sim 5\times 10^6$~K, centered around the magnetic pole.  
The bulk emission from the rest of the star is taken to be at a lower temperature and to contribute 
negligibly to the observed signal.  
We also assume that the size of the X-ray emission region is much smaller than the NS radius, $R$, with an 
approximately constant magnetic field normal to the NS surface.
The opening angular radius of the polar cap is defined to be $\beta < 2\pi$.
This geometry is consistent with what is observed in several magnetar sources.  
In quiescence and post-outburst, magnetars emit a thermal, soft X-ray component, with effective temperatures
$k_B T_{\rm eff}\sim 0.3-0.5$~keV (along with a non-thermal, hard X-ray power-law).  
Single and two-component blackbody fits to observed, thermal X-ray spectra infer 
emission radii $R_{\rm bb}\la 1$~km, 
implying that the radiation emerges from a small area \citep[e.g.,][]{Tiengoetal05a,Israeletal07a}.

Timing studies of AXPs and SGRs yield rotation periods of $P\sim 10$~s, and imply dipole magnetic 
field strengths $B\sim 10^{14}- 10^{15}$~G 
\citep[e.g.,][]{Kulkarnietal03a,Mereghettietal05b,Camiloetal07a,Israeletal07a}.\footnote{For a complete, 
current reference list of magnetar observations and measured quantities, see 
http://www.physics.mcgill.ca/$\sim$pulsar/magnetar/main.html}
\citet[][]{vanAdelsbergLai06a} showed that, for $P>0.01$~s, calculations of the phase-resolved photon Stokes 
parameters are independent of rotation period.

In \S\ref{subsect:Magnetar Atmospheres}, we briefly discuss radiative transfer in magnetar atmospheres.
In \S\ref{subsect:Vacuum Polarization}, we give a short review of the physics of vacuum polarization, 
and its effect on magnetar polarization.

\subsection{Magnetar Atmosphere Models}
\label{subsect:Magnetar Atmospheres}

Emission from magnetar surfaces is determined by radiative transfer through an atmosphere layer of ionized plasma 
\citep[see][for cases in which these conditions may be violated]{Hoetal03a,vanAdelsbergetal05a,MedinLai06a}.  
Due to gravitational setting, this atmosphere is likely to be composed of light elements, predominantly 
hydrogen or helium.

In a NS atmosphere, photons propagate in two polarization modes that have distinct interactions with the plasma medium. 
Ordinary (O) mode photons are linearly polarized, mostly in the plane formed by the photon propagation and magnetic field 
directions.  They have an absorption opacity approximately equal to that of photons in non-magnetic 
plasma.  Extraordinary (X) mode photons are 
linearly polarized mostly perpendicular to the plane formed by the propagation and magnetic field directions. 
They have an absorption opacity reduced by a factor 
$(E_{Be}/E)^2$, where $E_{Be} \approx 1158 B_{14}$~keV is the electron cyclotron energy, and $E$ is the photon 
energy \citep[e.g.,][]{Meszaros92a}.  The 
reduction factor in the X mode absorption opacity is due to the suppression of particle motion transverse to the strong 
magnetic field, and is as large as $10^6$ at photon energies $E\sim 1$~keV.  
X mode photons therefore decouple from deeper layers of the atmosphere than O mode photons, and, under typical conditions, 
dominate the emergent radiation.  Therefore, magnetar atmospheres emit strongly polarized radiation, with 
linear polarization fractions up to 100\% \citep[e.g.,][]{Pavlovetal94a,HoLai03a,LaiHo03b}.  
In addition, atmospheric emission 
is beamed along the magnetic field, though the beaming pattern broadens when the effects of vacuum 
polarization in strong magnetic fields are taken into account \citep[see][and the references therein]{vanAdelsbergLai06a}.

The atmosphere models of \citet[][]{vanAdelsbergLai06a} are the first to quantitatively incorporate the physics of 
vacuum polarization in strong magnetic fields, and 
produce accurate results for fully ionized atmospheres with field strengths $B = 10^{12}-10^{15}$~G.  
We use these radiative transfer calculations to construct the linear polarization fraction, defined:
\begin{eqnarray}
\Pi^{\rm em}_E = {F^O_E(\delta)-F^X_E(\delta)\over F^O_E(\delta)+F^X_E(\delta)},
\end{eqnarray}
where $F^O_E(\delta)$ and $F_E^X(\delta)$ are the observed intensities for X and O mode photons emerging 
from a hotspot on the star surface, and $\delta$ is the angle between the photon propagation direction and 
surface normal (see \S\ref{subsect:Emission from a Finite Spot}).

\subsection{Vacuum Polarization}
\label{subsect:Vacuum Polarization}

When the NS magnetic field strength is greater than the quantum critical field $B_Q\sim 4\times 10^{13}$~G 
(obtained by setting the electron cyclotron energy equal to its rest energy), there are significant vacuum contributions
to the 
dielectric tensor and magnetic permeability of the atmospheric plasma.  At certain photon energies, propagation angles, 
and matter densities, plasma and vacuum contributions cancel each other out, leading to a vacuum ``resonance'' in 
which both photon modes become circularly polarized and the approximation of geometric optics breaks down 
\citep[see][and the references therein]{Meszaros92a,LaiHo02a,HoLai03a}.  A photon of given energy $E$, propagating through 
the density 
gradient of the NS atmosphere, experiences two effects: (1) if it is X mode polarized, it experiences a significant 
enhancement of its absorption opacity in a narrow range of densities about the resonance; and (2) partial conversion 
occurs between X and O states in a manner analogous to the MSW effect for neutrinos \citep[][]{LaiHo02a}. 
The two primary effects of vacuum polarization on magnetar spectra are suppression of line features and softening of the 
hard tail present in previous models \citep[see, e.g.,][]{HoLai01a}.  
For details on the physics of vacuum polarization and its effect on radiative transfer in magnetar atmospheres, see 
\citet[][]{LaiHo02a}, \citet[][]{HoLai03a}, \citet[][]{LaiHo03b}, and \citet[][]{vanAdelsbergLai06a}.

It can be shown that photons of energy $E$, traveling through the vacuum resonance region, experience resonant mode 
conversion with probability $P_C = 1-\exp\left[-\pi(E/E_{\rm ad})^3/2\right]$.  
The adiabatic energy, $E_{\rm ad}$, is defined:
\begin{eqnarray}
E_{\rm ad}\approx 2.52 \left[ f_B \tan\theta_B\left|1-(E_{Bi}/E)^2\right|\right]^{2/3} H_{\rho}^{-1/3}\mbox{ keV},
\end{eqnarray}
where $\theta_B$ is the angle between the photon propagation direction and the magnetic field, $H_{\rho}$ is the 
atmosphere scale height, and $E_{Bi}\approx 0.63 (Z/A) B_{14}$~keV is the ion cyclotron energy, with atomic and mass 
numbers $Z$ and $A$, respectfully.  The parameter $f_B$ is a slowly varying function of $B$, whose magnitude is of order 
unity.  At photon energies $E\ll E_{\rm ad}$, mode conversion is ineffective, while for 
$E\ga 1.4 E_{\rm ad}$, mode conversion is essentially complete.  The vacuum resonance density occurs at:
\begin{eqnarray}
\rho_V\approx 0.96\, (Z/A)\, E_1^2\, B_{14}^2\, f_B^{-2}\ {\rm g\ cm}^{-3}.
\end{eqnarray}

The vacuum resonance phenomenon leaves a unique signature on the emission 
of polarized radiation from the NS \citep[][]{LaiHo03a}.  
For magnetic field strengths $B \ll 7\times 10^{13}$~G, photons of both modes 
encounter the vacuum resonance after decoupling from the NS atmosphere, while for 
$7\times 10^{13}\mbox{ G}<B< 5\times 10^{16}$~G, the vacuum resonance occurs at an atmospheric depth between the X and O 
mode photospheres \citep[][]{LaiHo03a}.  
In the former case, photons encounter the vacuum resonance while propagating into the magnetosphere.
For low energy photons with $E\ll E_{\rm ad}$, 
mode conversion is ineffective, and the vacuum resonance has no effect on the emerging radiation.  
For high energy photons, mode conversion is effective, and results in conversion between the X and O polarization 
states. 
Thus, the vacuum polarization effect causes the plane of linear polarization to rotate $90^{\circ}$ between 
low and high energies.  This can occur in models with magnetic fields as high as $B=7\times 10^{13}$~G, though there is 
a dependence on the NS geometry (see \S\ref{sect:Results}).

\section{Emission Model}
\label{sect:Emission Model}

\subsection{Evolution of Photon Stokes Parameters in a Dipole Magnetosphere}
\label{subsect:Observed Polarization}

After emerging from the atmosphere, photons propagate through the NS magnetosphere before reaching the 
observer.  As discussed above, we ignore scattering and only consider magnetic field effects on the polarization mode 
evolution.

The solution to Maxwell's equations in magnetized vacuum yields the 
unit polarization vectors for X and O mode photons in the magnetosphere
\citep[e.g.,][]{vanAdelsbergLai06a}:
\begin{eqnarray}
{\bf\hat{e}}_O & = & (\cos\varphi_B,\sin\varphi_B),\\
{\bf\hat e}_X & = & (-\sin\varphi_B,\cos\varphi_B),
\end{eqnarray}
where $\varphi_B$ is the azimuthal angle of the magnetic field projected in a plane perpendicular to the observer line of 
sight (see below).  
Thus, the photon polarization state depends only on the magnetic field direction.  The difference in 
the eigenvalues corresponding to the X and O polarization states is 
$E \Delta n/(\hbar c)\sim 5\times 10^3\, E_1\, B_{14}^2\, (r/R)^{-6}$~cm$^{-1}$, where $E_1 = E/(\mbox{1 keV})$ is the 
photon energy at the NS surface and $r$ is the distance between the photon and the NS.
When $r\ga R$, the derivative of the angle for a dipole field is approximately 
$d \varphi_B/ds \sim r^{-1} = 10^{-6}\, r_{10}^{-1}$~cm$^{-1}$, 
where $r_{10} = r/(10\mbox{ km})$ and $s$ is the affine parameter defined along the photon geodesic.  
Thus, near the star surface, $E\Delta n/(\hbar c)\gg d\varphi_B/ds$, and the polarization state evolves adiabatically 
with the 
changing direction of the magnetic field.  As the photon continues to propagate in the magnetosphere, it reaches the 
polarization limiting radius, $r_{\rm pl}$, where $E\Delta n/(\hbar c) = d\varphi_B/ds$.  
At distances greater than $r_{\rm pl}$, 
the polarization state is fixed, and the directions of the mode vectors are constant.  Thus, the 
measured values of the photon Stokes parameters are ``frozen in'' at $r_{\rm pl}$.  Far from the NS, 
$d\varphi_B/ds\sim 1/r_l$, where $r_l\equiv c/\Omega$ is the light-cylinder radius. 
Therefore, $r_{\rm pl}$ takes the value:
\begin{eqnarray}
r_{\rm pl}\sim 1.5\times 10^3\left({E_1 B_{14}^2 f_1^{-1}}\right)^{1/6} R_{10} \mbox{ km}.
\end{eqnarray}

In contrast, if the polarization states of emitted photons were determined near the NS surface, the observed signal would 
be greatly reduced.  In this case, addition of the Stokes parameters for photons emitted from regions with distinct 
magnetic field directions would tend to cancel.  Adiabatic evolution of the photon modes to distances far from the 
star surface, where the magnetic geometry is uniform, leads to significant polarization signals, even when emission 
occurs over an extended region on the star surface \citep[][]{HeylShaviv02a,Heyletal03a}.

We present calculations for NSs with a pure dipole magnetic field structure.  As discussed above, realistic magnetar 
models include twisted magnetic field configurations which contain contributions from higher-order multipoles.  
Nevertheless, we have argued that because the polarization signal is fixed far from the star surface, the dipole component 
of the field is the most important for determining the observed Stokes parameters.  \citet[][]{WangLai09a} performed 
a detailed analysis of polarization mode evolution in a strongly magnetized vacuum, and showed that adiabaticity can be 
broken near the star surface only when the photon traverses a quasi-tangential region (where the photon direction is along 
the magnetic field).  They presented a semi-analytic formalism for incorporating this effect into future 
calculations; however, for the polar cap sizes considered in this paper, $\beta = 5 - 30^{\circ}$, the correction will be 
small.

\citet[][]{LaiHo03a} used the adiabaticity of photon mode evolution in the magnetosphere to develop a simple formalism 
for calculating the observed photon Stokes parameters.  The results of their calculation are accurate for NSs with 
rotation periods $P>0.01$~s \citep[][]{vanAdelsbergLai06a}.
Following the work of \citet[][]{LaiHo03a}, we employ a fixed coordinate system, $xyz$, in which the observer line of sight 
is along the $z$ axis.  The rotation axis, {\boldmath{$\hat{\Omega}$}} is defined to be in the $xz$ plane, such that 
{\boldmath{$\hat{\Omega}$}}$\times {\bf\hat{z}} = \sin\alpha_R\, {\bf\hat{y}}$, where $\alpha_R$ is the angle between 
the rotation and $z$ axes.  The NS magnetic dipole moment vector, {\boldmath{$\mu$}}, 
is inclined at an angle $\alpha_M$ relative to the rotation axis.

We define the rotational phase $\psi$ as the azimuthal angle subtended by the magnetic dipole vector around the axis of 
rotation.  The phase is taken to be $\psi=0$ when {\boldmath{$\hat{\mu}$}} is in the $xz$ plane.
The angle between 
{\boldmath{$\mu$}} and $z$ is given by:
\begin{eqnarray}
\label{eq:cosTheta}
\cos\Theta = \cos\alpha_R\cos\alpha_M+\sin\alpha_R\sin\alpha_M\cos\psi.
\end{eqnarray}

The NS magnetic field, in the near-zone of the star such that $r\ll \Omega/c$, can be calculated according to the standard 
dipole formula, ${\bf B} = \left[3(\mbox{\boldmath{$\mu$}}\cdot{\bf\hat r}){\bf\hat r}-\mbox{\boldmath{$\mu$}}\right]/r^3$.
Along the observer line of sight, $r = z$, and the projection of ${\bf B}$ into the $xy$ plane yields:
\begin{eqnarray}
\cos\varphi_B & = & \left(\sin\alpha_R\cos\alpha_M-\cos\alpha_R\sin\alpha_M\cos\psi\right)/|\sin\Theta|,\\
\sin\varphi_B & = &  -\sin\alpha_M\sin\psi/|\sin\Theta|.
\end{eqnarray} 
Far from the NS, the photon geodesic is approximately 
${\bf r} \approx r{\bf\hat{z}}$.  The phase can be written $\psi = \psi_{\rm em} + r_{\rm pl}/r_l$, where $\psi_{\rm em}$ 
is the value of the rotational phase when the photon is emitted.  
For magnetars, which rotate slowly, $r_{\rm pl}/r_l\ll 1$, and $\psi\approx\psi_{\rm em}$.

For rapidly rotating NSs with $f_1\ga 100$, an accurate calculation of the observed polarization requires 
integration of the transfer equations for the photon Stokes parameters in the magnetosphere, which can result in 
significant circular polarization
\citep[][]{vanAdelsbergLai06a}.  Circular polarization can also be generated in cases where the polarization 
limiting radius occurs close to either the NS surface or light cylinder radius \citep[][]{HeylShaviv02a}.
However, for magnetars with adiabatic evolution of the modes, the normalized Stokes parameters take the simple form:
\begin{eqnarray}
\label{eq:Q_E}
Q_E/I_E & = & \Pi^{\rm em}_E\cos\left(2\varphi_B\right),\\
\label{eq:U_E}
U_E/I_E & = & \Pi^{\rm em}_E\sin\left(2\varphi_B\right),
\end{eqnarray}
where $I_E\equiv I^O_E(\theta_{\rm em})+I^X_E(\theta_{\rm em})$ is the total specific intensity.
The $Q_E$, $U_E$ Stokes parameters define the plane of polarization in the $xyz$ coordinate sytem.  Thus, measurements 
of the photon polarization directly map out the rotation of the magnetic dipole field around the NS. 

\subsection{Emission from a Finite Spot}
\label{subsect:Emission from a Finite Spot}

Phase-dependent emission from an extended area on a NS surface is computed according to the method described by 
\citet[][]{Pechenicketal83a}, using the generalizations of \citet[][]{PernaGotthelf08a}. 
We assume that the radiation emerges from a single spot, centered around the magnetic dipole vector, with opening 
angle $\beta$.  
Thus, the angle between the spot center and observer line of sight is given by equation~(\ref{eq:cosTheta}).
We describe points on the NS surface using the polar angle $\theta$ and azimuthal angle $\varphi$, in spherical 
polar coordinates.  If $\Theta=0$, $\theta$ is restricted to $\theta\le\beta$, otherwise the condition becomes:
\begin{eqnarray}
\Theta-\beta & \le \theta \le & \Theta+\beta\\
2\pi-\varphi_{\star} & \le \varphi \le & \varphi_{\star},
\end{eqnarray}
where
\begin{eqnarray}
\cos\varphi_{\star} = {\cos\beta-\cos\Theta\cos\theta\over\sin\Theta\sin\theta}.
\end{eqnarray}
When $0<\Theta<\beta$, the constraints on $\theta$ and $\varphi$ become:
\begin{eqnarray}
\theta & \le & \theta_{\star}(\Theta,\varphi,\beta),\\
\label{eq:numeric}
\cos\beta & = & \sin\Theta\sin\theta_{\star}\cos\varphi+\cos\Theta\cos\theta_{\star},
\end{eqnarray}
where equation~(\ref{eq:numeric}) must be solved numerically.  The values of $\theta$ and $\varphi$ that describe points 
in the spot are 
restricted to those that yield solutions for co-angles $\varphi_{\star}(\beta,\Theta,\theta)$ 
or $\theta_{\star}(\beta,\Theta,\varphi)$, depending on the condtions. 

Due to general relativistic effects, a photon emitted from colatitude $\theta$ will reach 
the observer if emitted at an angle $\delta$ with respect to the
surface normal, where the relation between the two angles is given by
the ray-tracing function \citep[][]{Pechenicketal83a,Page95a}:
\begin{eqnarray}
\theta(\delta) = \int_0^{R_s/2R}\,du\, x\,\left[\left(1-{R_s\over R}\right)\left({R_s\over 2R}\right)^2-
	(1-2u)u^2 x^2\right]^{-1/2}.
\label{eq:teta}
\end{eqnarray} 
In equation~(\ref{eq:teta}), $x\equiv\sin\delta$, $R_s\equiv 2GM/c^2$ is the Schwarzchild radius, and $M$ is the 
NS mass.  In our calculations, we assume that $M=1.4 M_\odot$ and $R=12$ km. 
A simpler, more convenient relationship between the emission angle and colatitude is given by the approximation in 
\citet[][]{Beloborodov02a}:
\begin{eqnarray}
\cos\delta\approx  {R_s\over R} + \left(1-{R_s\over R}\right)\cos\theta.
\label{eq:tetapprox}
\end{eqnarray}
Equation~(\ref{eq:tetapprox}) differs from (\ref{eq:teta}) by less than 1\% when $R>3R_s$.  The portion of the NS surface 
visible to the observer is given by the set of $\theta$ values that yield solutions for $\delta$ in 
equation~(\ref{eq:tetapprox}).

The total, phase-dependent specific intensities from the spot are then obtained by integrating the local emission 
in each mode over the observable surface, accounting for the gravitational redishift of the 
radiation \citep[][]{Page95a}:
\begin{eqnarray}
F^j(E_{\infty},\psi) = \int_0^1\,dx\, x\,\int_0^{2\pi}\, d\varphi\, I_j\left(\theta,\varphi,E_{\infty}e^{-\Lambda}\right),
\end{eqnarray}
where the local specific intensity in each mode $I_j$ ($j=\mbox{X, O}$) is set to zero outside of the 
boundaries on $\theta$ and $\varphi$ set above.  The energy observed at infinity is given by 
$E_{\infty}=E e^{\Lambda}$, with
\begin{eqnarray}
e^{\Lambda} = \sqrt{1-{R_s\over R}}.
\end{eqnarray}
The phase-dependent Stokes parameters are then readily calculated using the methods of \S\ref{subsect:Observed Polarization}.

\section{Results}
\label{sect:Results}

We present results from our magnetar models with fully ionized hydrogen atmospheres, magnetic field strengths 
$7 \times 10^{13}$~G -- $5\times 10^{14}$~G, and effective temperature $T_{\rm eff}=5\times 10^6$~K (a value which is in the
typical magnetar range).   
We explore four representative geometries: $\alpha_R=90^{\circ}$, $\alpha_M=90^{\circ}$ (hereafter denoted G1); 
$\alpha_R=90^{\circ}$, $\alpha_M=45^{\circ}$ (hereafter denoted G2); $\alpha_R=45^{\circ}$, 
$\alpha_M=45^{\circ}$ (hereafter denoted G3); and $\alpha_R=45^{\circ}$, $\alpha_M=0^{\circ}$ (hereafter denoted G4).

\subsection{Atmosphere Model Linear Polarization Fractions}
\label{subsect:EmPolFrac}

As discussed above, quiescent magnetar emission spectra are well fit by a blackbody plus power-law
and do not contain absorption features.  Thus, 
it is difficult to distinguish magnetic atmosphere models from phenomenological fits.  However, the linear polarization 
is strongly dependent on the magnetic field and viewing geometry, and, importantly, cannot be predicted without a detailed 
physical model for the surface emission.  In this section, we describe the dependence of $\Pi_{\rm em}$ on magnetic field 
and geometry.

\begin{figure}
\centering
\includegraphics[scale=0.25]{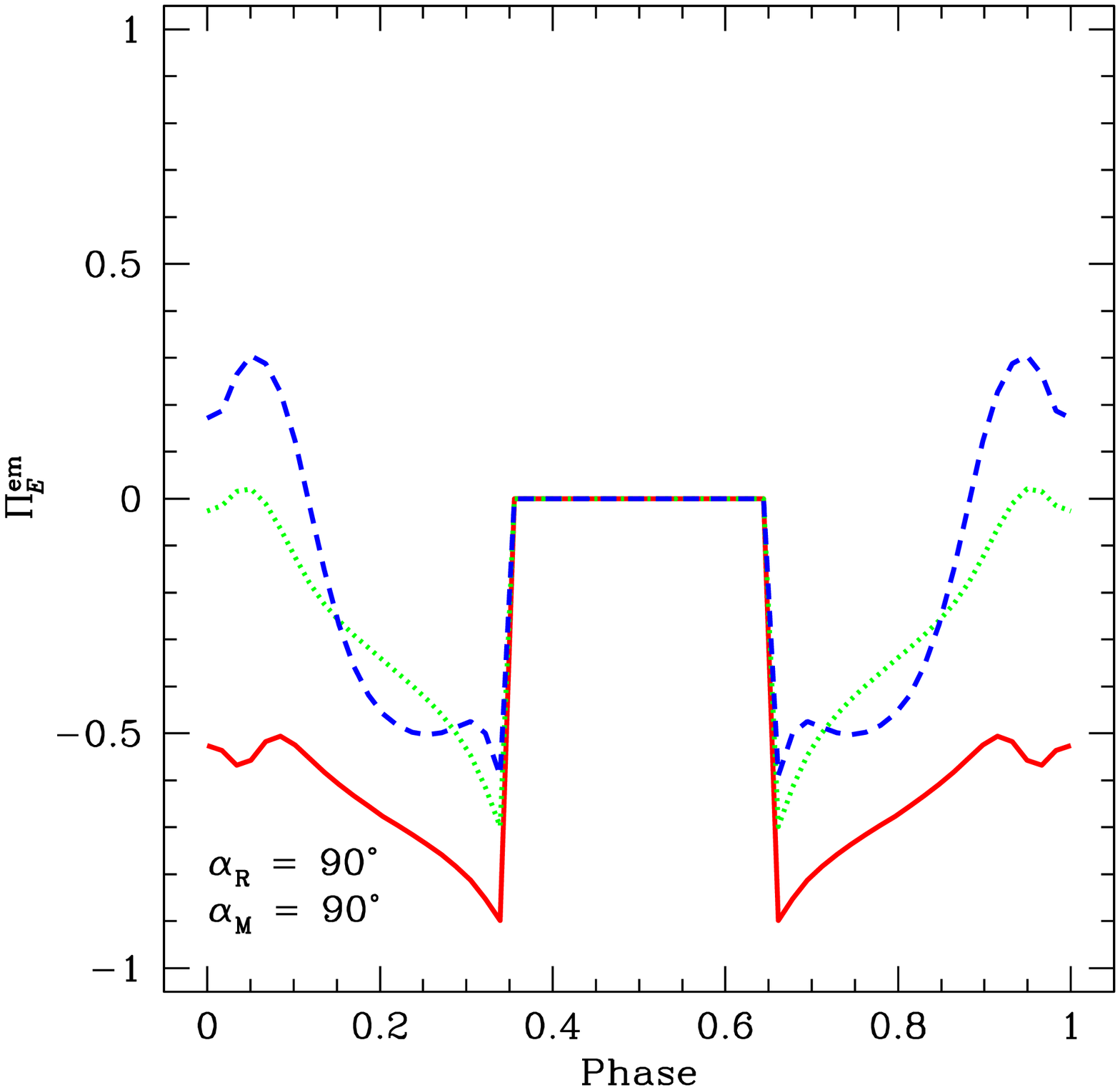}
\includegraphics[scale=0.25]{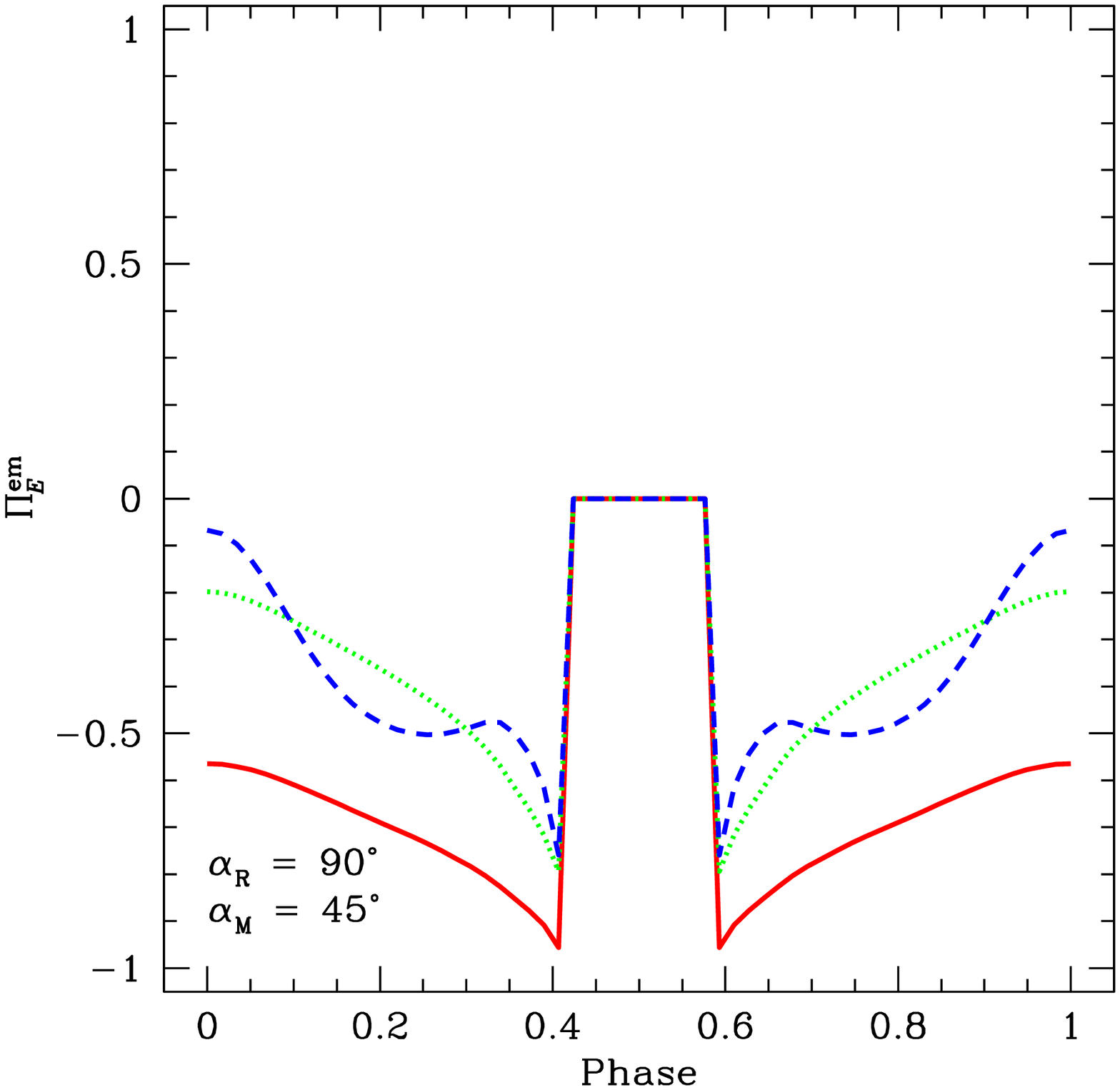}\\
\includegraphics[scale=0.25]{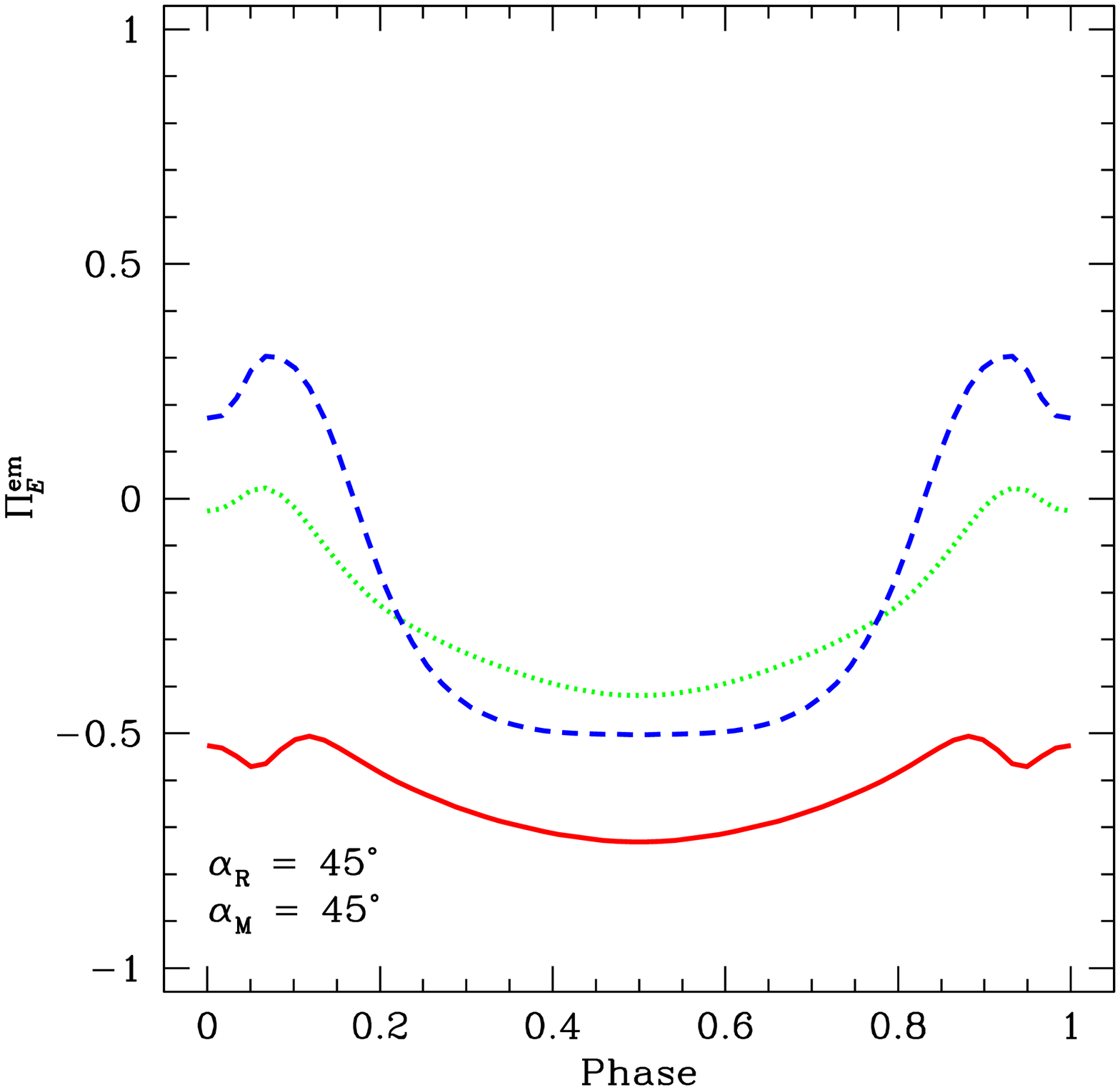}
\includegraphics[scale=0.25]{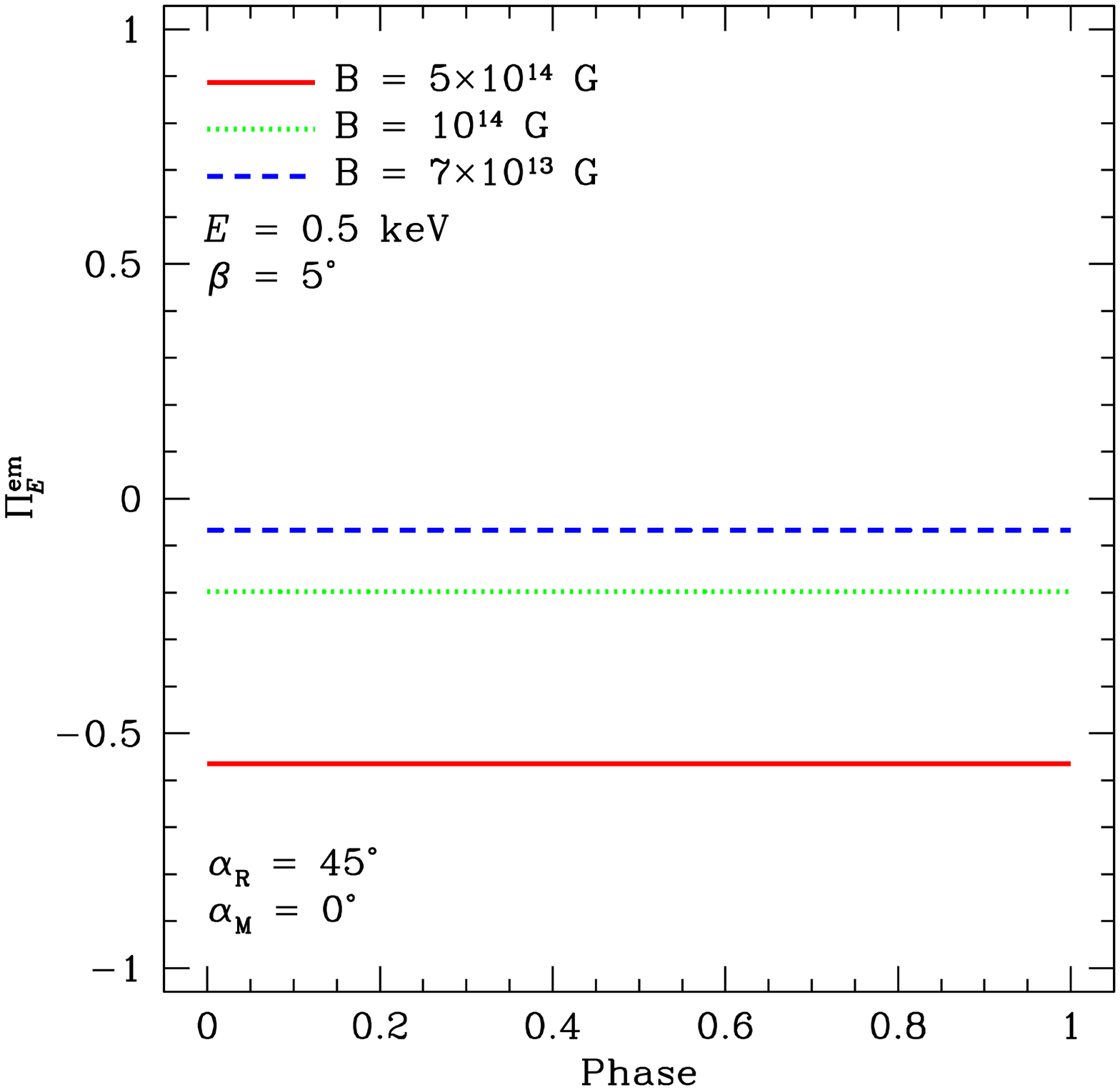}
\caption{Emitted polarization fraction, $\Pi^{\rm em}_E\equiv (I_O-I_X)/(I_O+I_X)$, as a function of phase, 
for an extended spot on a NS surface.  
The spot is located at the magnetic pole and has opening angle $\beta = 5^{\circ}$.
Results are shown for four NS geometries and three magnetic 
field strengths: $B=5\times 10^{14}$~G (solid curve), $B=10^{14}$~G (dotted curve), and $B=7\times 10^{13}$~G 
(dashed curve).  The angles $\alpha_R$ and $\alpha_M$ are the polar inclinations of the rotation axis relative to the 
line of sight, and the magnetic dipole vector relative to the rotation axis, respectively.  The photon energy is set to 
$E=0.5$~keV.}
\label{fig:pfracE5}
\end{figure}

\begin{figure}
\centering
\includegraphics[scale=0.25]{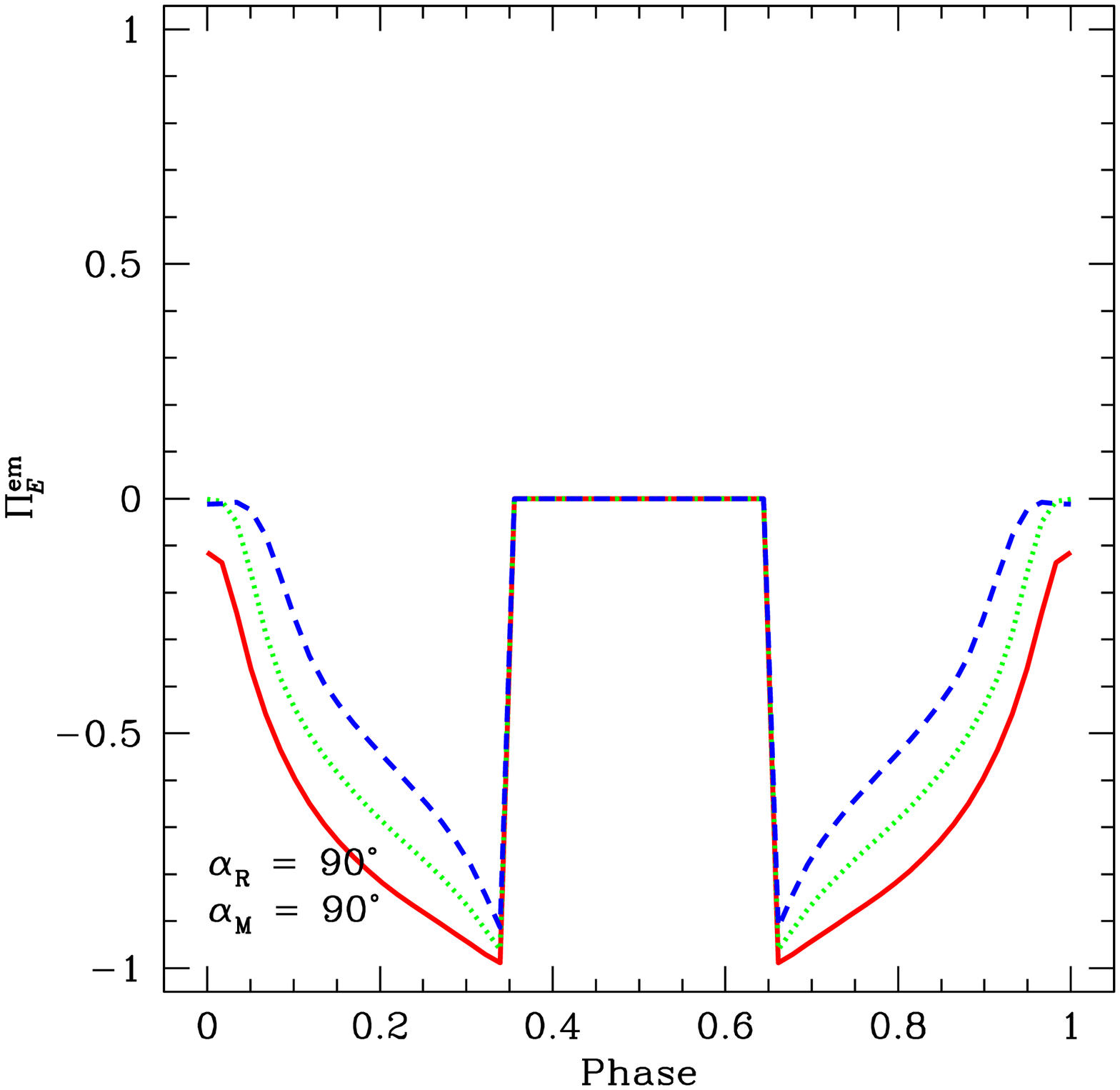}
\includegraphics[scale=0.25]{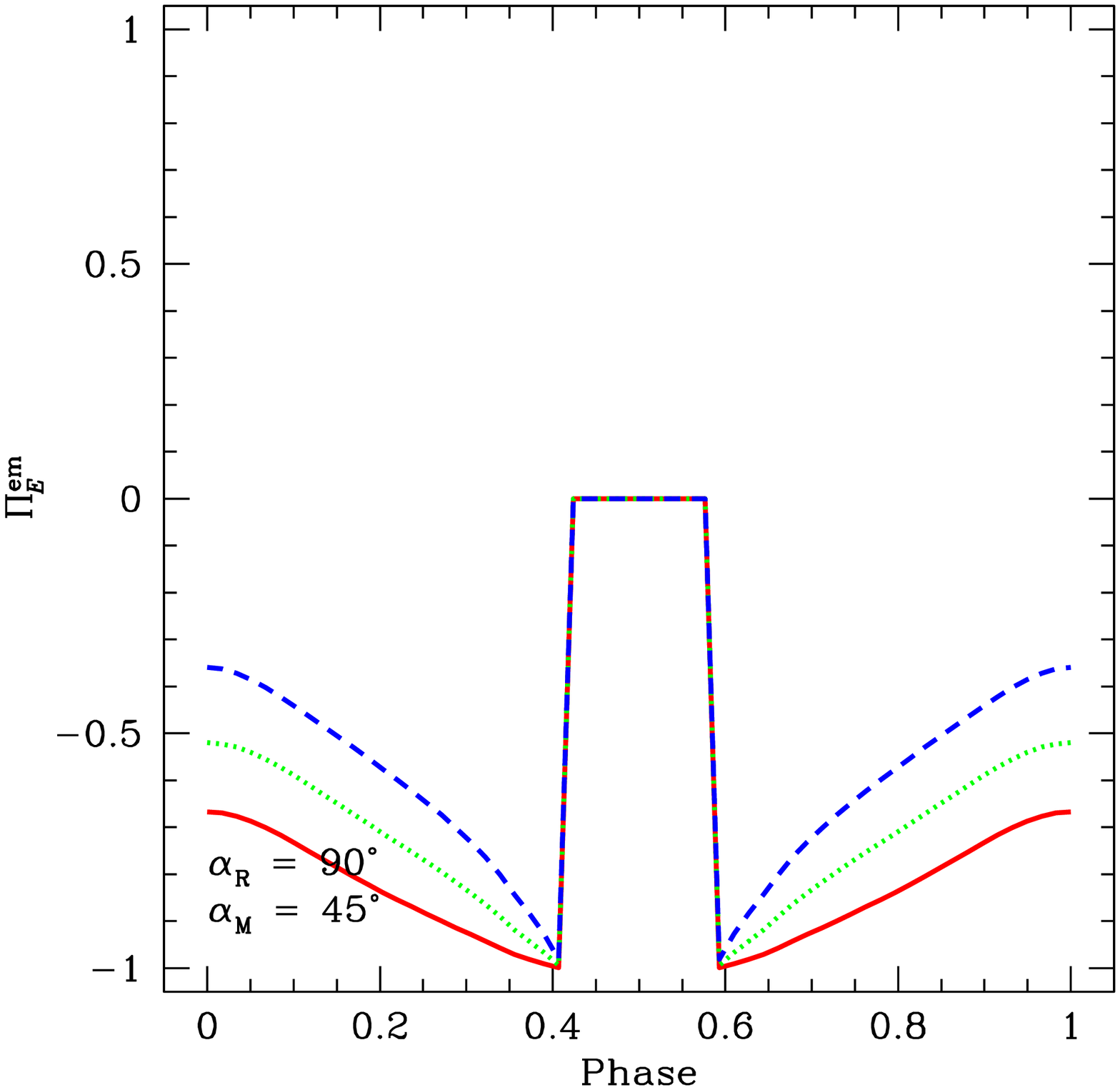}\\
\includegraphics[scale=0.25]{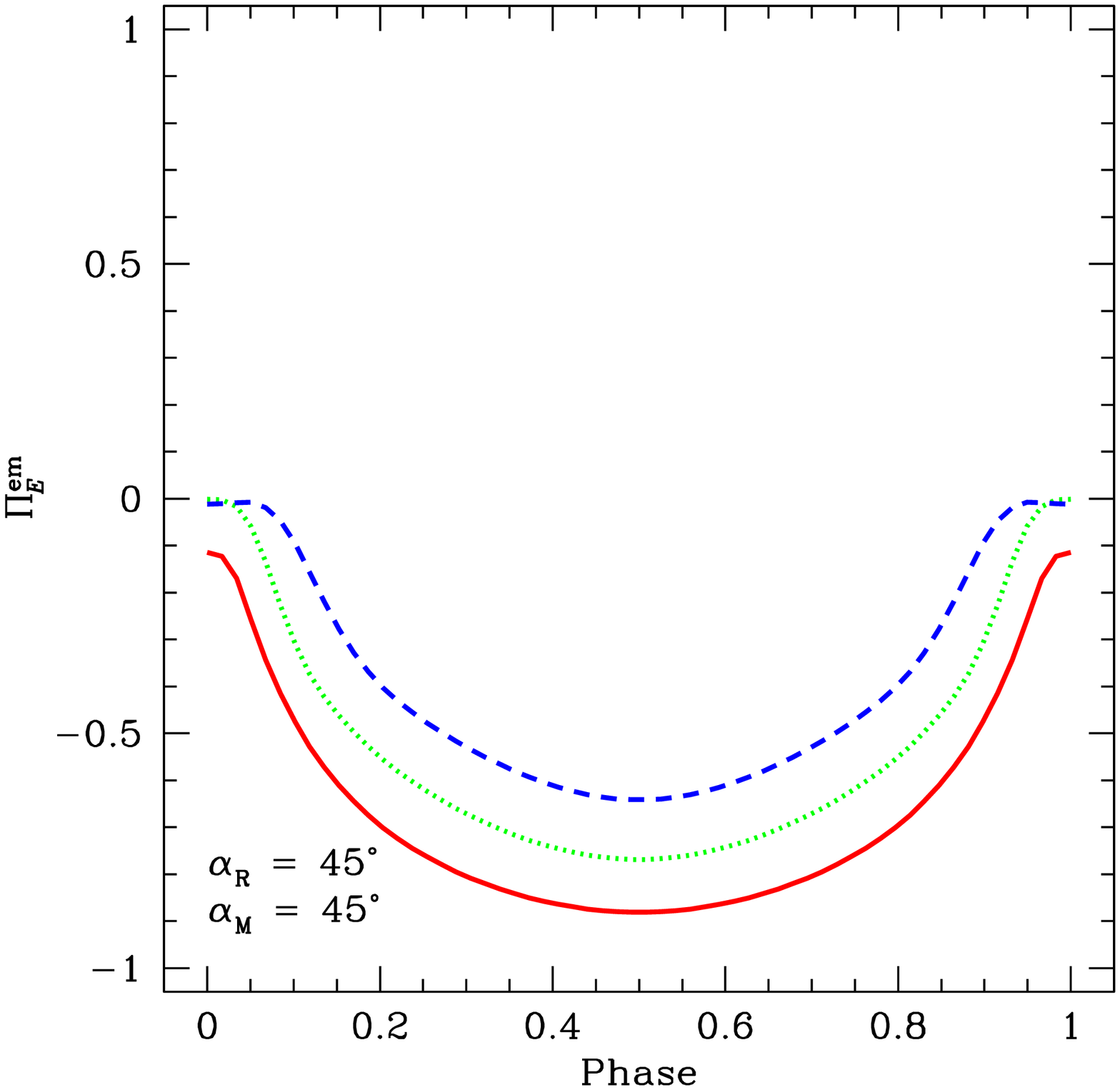}
\includegraphics[scale=0.25]{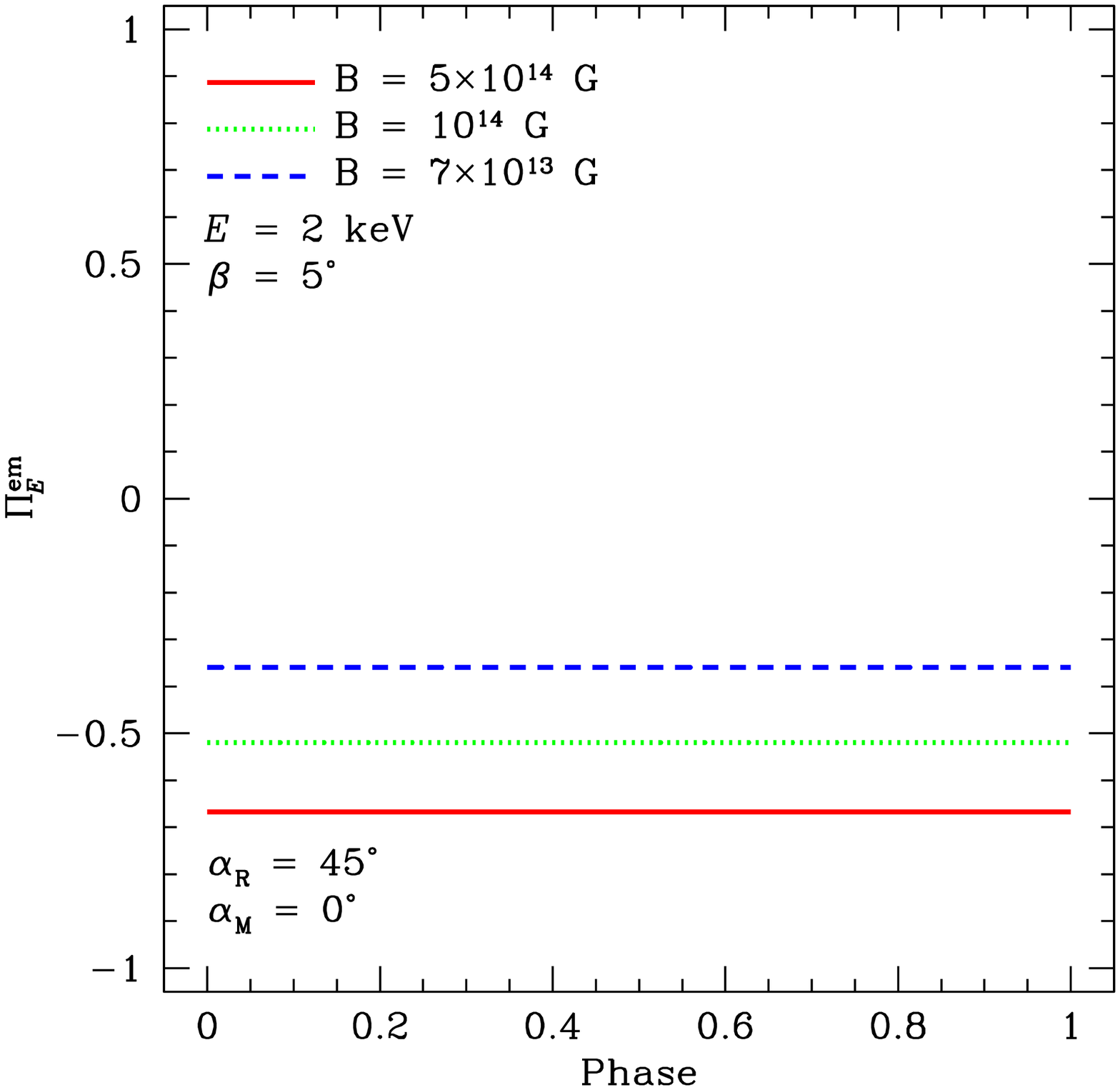}
\caption{Emitted polarization fraction, $\Pi^{\rm em}_E\equiv (I_O-I_X)/(I_O+I_X)$, as a function of phase, 
for an extended spot on a NS surface.  
The spot is located at the magnetic pole and has opening angle $\beta = 5^{\circ}$.
Results are shown for four NS geometries and three magnetic 
field strengths: $B=5\times 10^{14}$~G (solid curve), $B=10^{14}$~G (dotted curve), and $B=7\times 10^{13}$~G 
(dashed curve).  The angles $\alpha_R$ and $\alpha_M$ are the polar inclinations of the rotation axis relative to the 
line of sight, and the magnetic dipole vector relative to the rotation axis, respectively.  The photon energy is set to 
$E=2$~keV.}
\label{fig:pfracE20}
\end{figure}

Figures~\ref{fig:pfracE5} and \ref{fig:pfracE20} show the linear polarization fraction calculated using NS 
atmosphere models at magnetic field strengths $B=5\times 10^{14}$~G (solid curves), 
$B=10^{14}$~G (dotted curves), and $B=7\times 10^{13}$~G (dashed curves).  Each panel in the figure corresponds to one of 
the NS geometries described above.  The polar cap size is set to $\beta = 5^{\circ}$.

Figure~\ref{fig:pfracE5} shows results for photons with energy $E=0.5$~keV.  The upper-left panel depicts the case of 
the ``orthogonal rotator,'' in which the magnetic 
dipole vector rotates in the $yz$ plane, intersecting the line of sight at $\psi = 0$.  The upper-right and lower-left 
panels show less extreme geometries.  In G2, most of the photons are emitted with $\delta\ga \pi/4$, and in G3, the dipole 
vector intersects the line of sight when $\psi = 0$, but sweeps out a cone whose base is perpendicular to the NS 
rotation axis.  The lower-right panel shows the extreme case when the rotational and magnetic axes are aligned.

There are two key features of Figure~\ref{fig:pfracE5} that highlight the interplay between the NS geometry, magnetic 
field strength, and vacuum polarization effects to produce the emitted polarization.  These features include:
(1) $|\Pi^{\rm em}|$ is 
smaller for values of $\psi$ close to $0$ than for values of $\psi$ close to $\pi/2$; 
(2) at $\psi\approx 0$, the sign of $\Pi^{\rm em}$ is positive for $B=7\times 10^{13}$~G, and negative 
for stronger magnetic fields.

Result (1) can be understood by considering the variation in X and O mode opacities with angle between the photon 
propagation and magnetic field directions.  For the orthogonal rotator, small phases roughly correspond to emission 
along the observer line of sight and magnetic field direction.
For emission angles $0<\delta\ll 2\pi$, the difference between the free-free absorption opacities of the X and O modes 
is smaller than for angles $\delta\approx \pi/4$
\citep[see \S2.6 of][]{HoLai01a}.  The densities at which X and O mode photons decouple from the atmosphere are therefore
closer in value, leading to the smaller magnitude of $\Pi^{\rm em}$.  Conversely, at phases for 
which $\psi\ga \pi/2$, emission angles for photons reaching the observer are $\delta\la \pi/4$.  
At these angles, the difference between the X and O mode opacities 
(and hence, decoupling depths) is maximal, yielding larger values of $|\Pi^{\rm em}|$.  
This trend is also visible in geometries
G2 and G3 of Figure~\ref{fig:pfracE5}, where  
in the G3 case, emission angles $\delta\approx \pi/4$ are realized for
phases $\psi\approx \pi$. 
In the G4 geometry, the emission angle and hence the polarization are independent of the phase.

Result (2) is a manifestation of the vacuum polarization effect.  As
discussed above, for typical angles and photon energies, the emitted
specific intensity of X mode photons is greater than that of O mode
photons.  Under these conditions, $\Pi^{\rm em}<0$.  
However, at magnetic field strength $B=7\times 10^{13}$~G, photons with energy
$E=0.5$~keV encounter the vacuum resonance at a density $\rho_V\la
\rho_O$, where $\rho_O$ denotes the O mode decoupling density.  Thus,
if resonant mode conversion occurs, $\Pi^{\rm em}$ will switch sign,
corresponding to a rotation of the plane of linear polarization.  
At emission angles $\delta\approx \pi/4$ and photon energies $E=0.5$~keV, the
mode conversion probability is much less than unity.  However, when the emission angle
$\delta\approx 0$, the mode evolution becomes adiabatic, and $P_C\rightarrow
1$.  Therefore, at phases for which the emission angle is small, resonant
mode conversion occurs and the sign of $\Pi^{\rm em}$ changes.  For
magnetic fields $B > 10^{14}$~G, the vacuum resonance density is 
intermediate between the X and O mode decoupling densities, and no change in the plane
of linear polarization occurs.  Rotation of the plane of linear polarization at $B < 7\times 10^{13}$~G occurs 
in geometries G1 and G3 (upper- and lower-left panels), in which $\psi\rightarrow 0$ corresponds to 
$\delta\rightarrow 0$.  
In geometries G2 and G4, (upper- and lower-right panels), the 
emission occurs at angles $\delta\approx \pi/4$, resulting in little mode conversion at 
$E = 0.5$~keV.

As the magnetic field strength is increased, the suppression factor in the X
mode free-free absorption opacity increases as $B_{14}^2$.  Thus, we
expect the emitted polarization fraction to increase (roughly) with the strength
of the atmosphere magnetic field.  When vacuum effects are taken into
account, the decoupling depth of X mode photons is effectively reduced
by the increase in free-free absorption opacity and mode
conversion at the resonance \citep[][]{HoLai03a}.  If the X mode photon
decoupling density is approximately equal to the vacuum resonance
density, the trend is preserved ($\rho_V\propto B_{14}^2$).  This results in large 
values of $|\Pi^{\rm em}|$ at $B=5\times 10^{14}$~G relative to those 
at $B=7\times 10^{13}$~G.

Figure~\ref{fig:pfracE20} shows the same results as
figure~\ref{fig:pfracE5}, except for photons with $E=2$~keV.  While
the general variation in $|\Pi^{\rm em}|$ with phase remains the same
as in the $E=0.5$~keV case, the rotation of the plane of polarization 
at $\psi\approx 0$ no longer occurs.  This is because, at $B=7\times
10^{13}$~G, photons with $E=2$~keV encounter the vacuum resonance at a
larger density, which occurs between the X and O mode photospheres for 
$\delta\ll 1$, but at a 
smaller density than photons in models with $B=10^{14}$~G.  Thus,
there is no rotation of the plane of linear polarization at the
lower field strength, and the magnitude of the polarization fraction
$|\Pi^{\rm em}|$ is larger for $B=10^{14}$~G.

\subsection{Observed Stokes Parameters}

The observed polarization signals are calculated by multiplying the
atmosphere polarization fraction $\Pi^{\rm em}_E$ by the angular
factors in equations~(\ref{eq:Q_E}) and (\ref{eq:U_E}).  Thus, the
characteristics of intrinsic atmosphere emission are modulated by
the projection of the NS magnetic field direction into the $xy$ plane.

\begin{figure}
\centering
\includegraphics[scale=0.25]{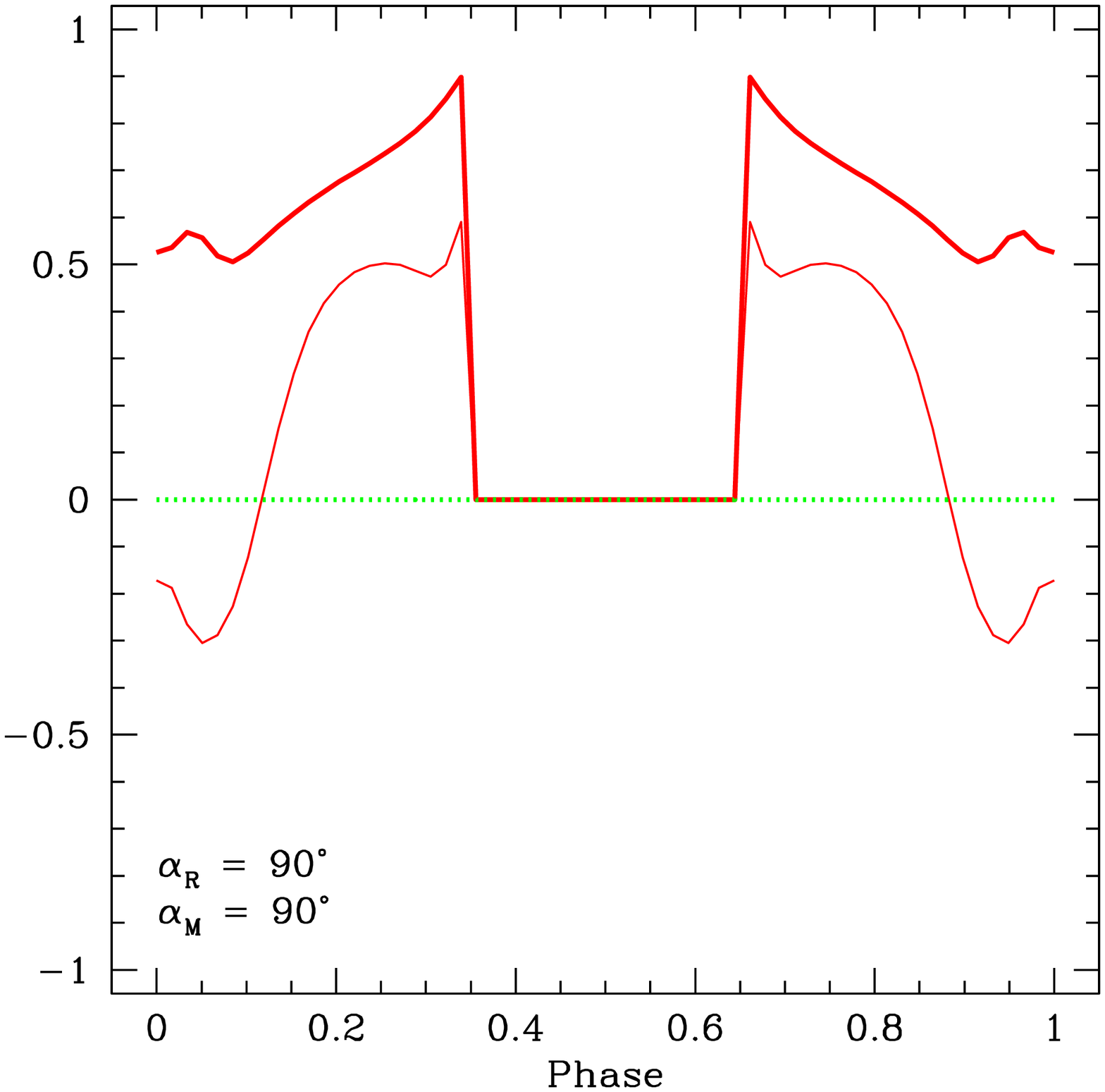}
\includegraphics[scale=0.25]{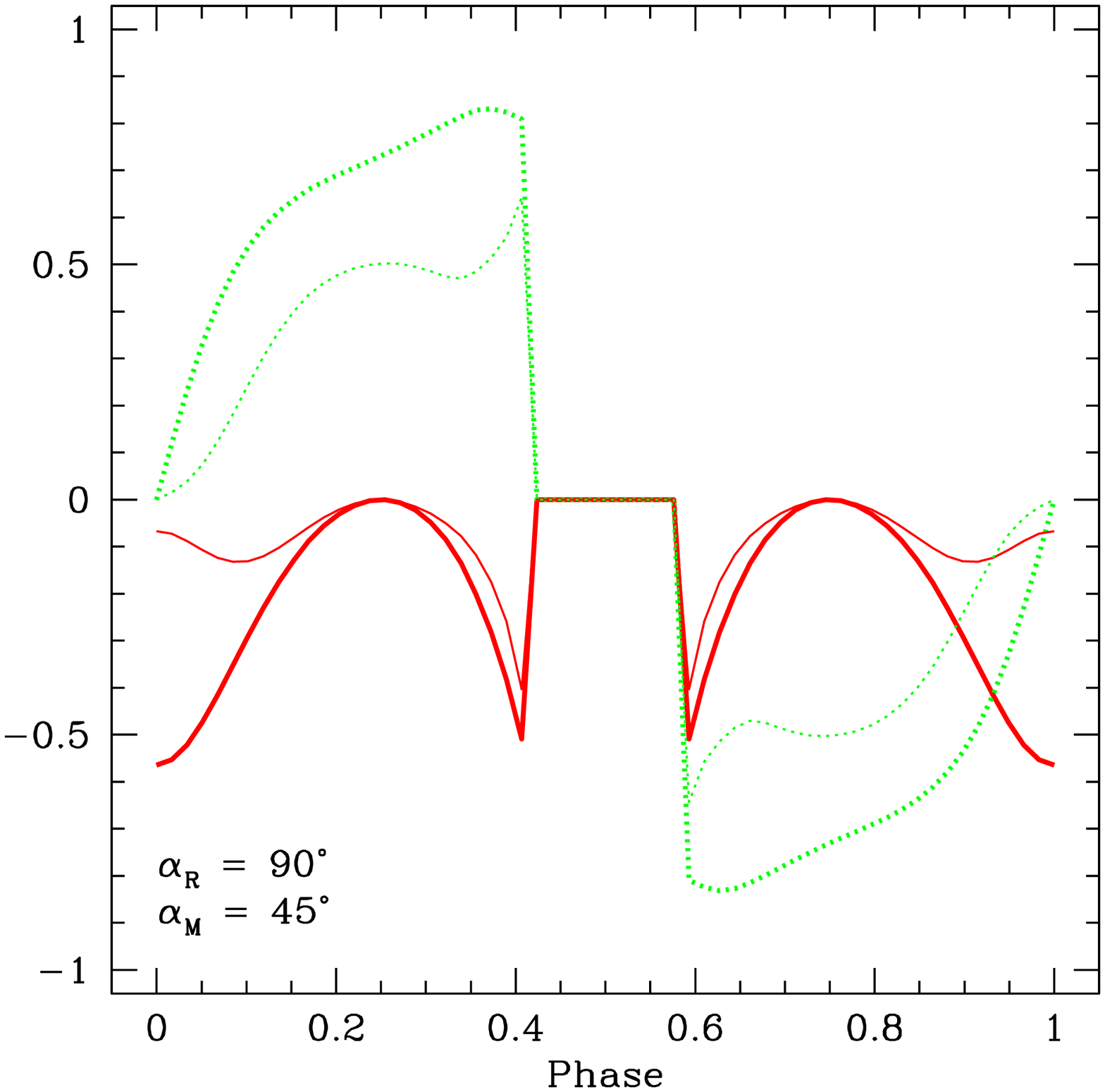}\\
\includegraphics[scale=0.25]{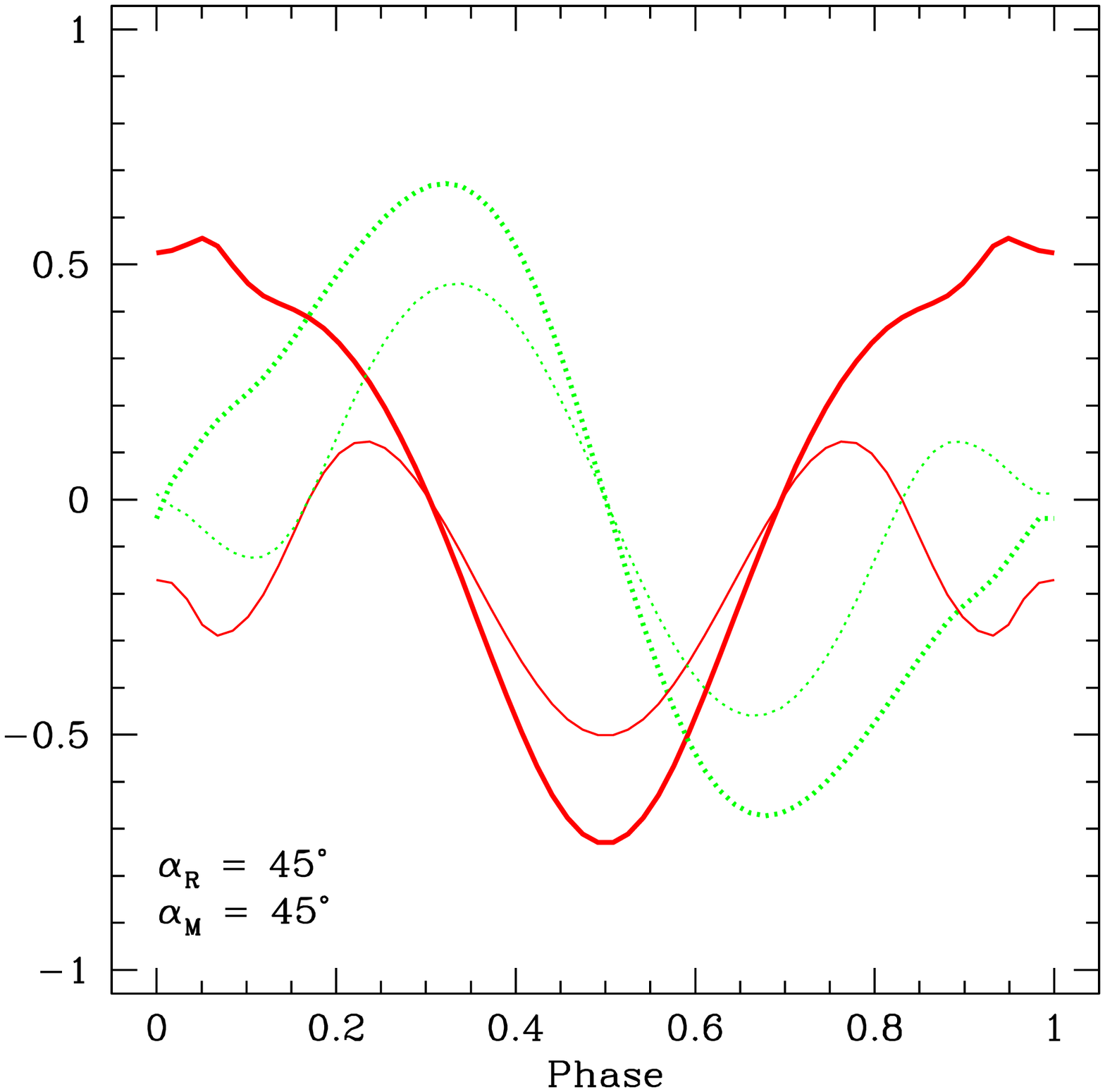}
\includegraphics[scale=0.25]{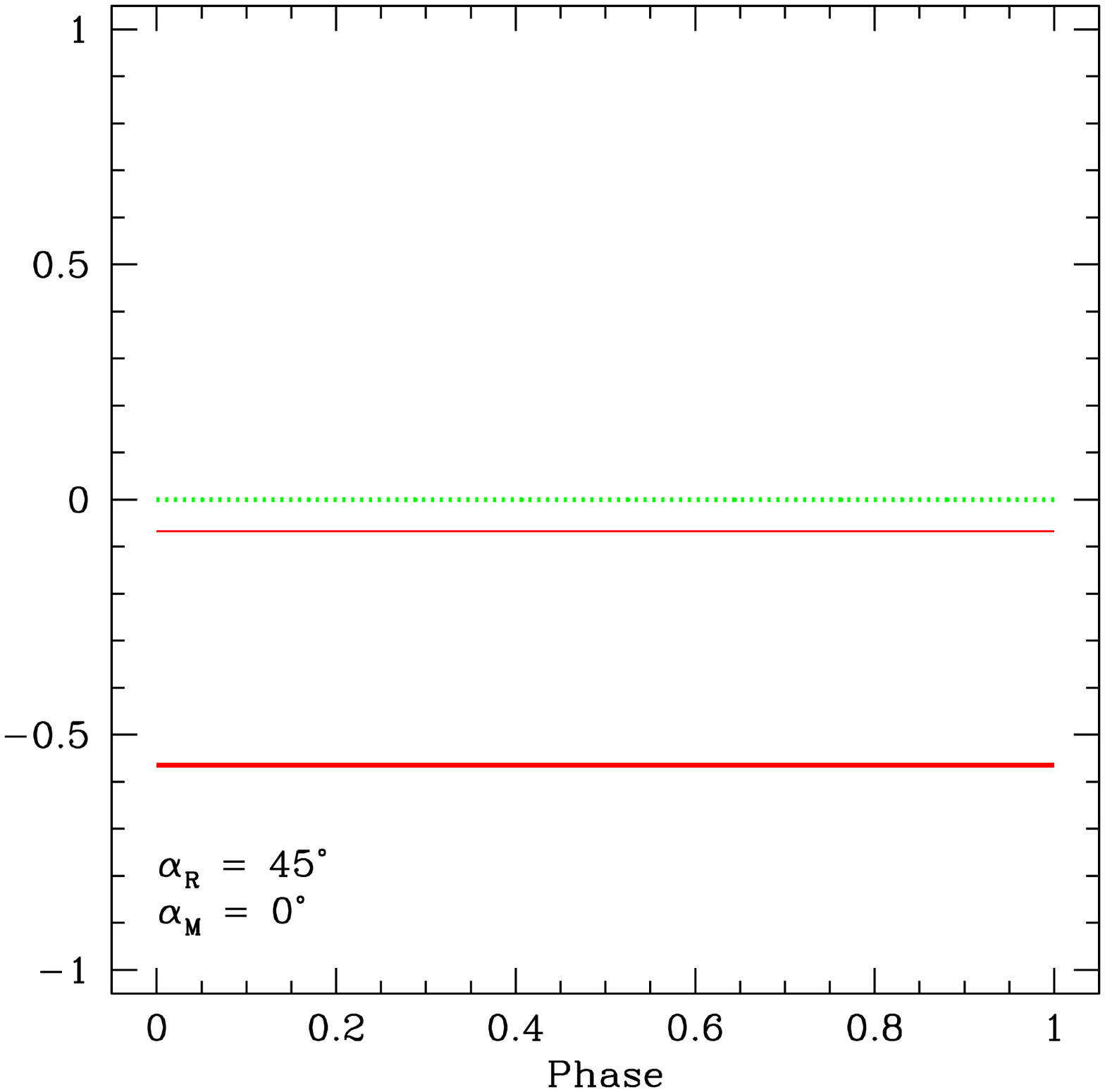}
\caption{Observed, normalized Stokes parameters $Q/I$ (solid curves), $U/I$ (dotted curves), as a function of phase, 
for an extended spot on a NS 
surface.  The spot has opening angle $\beta = 5^{\circ}$.  Results are shown for the same geometries as in 
figure~\ref{fig:pfracE5}, for magnetic field strengths 
$B=5\times 10^{14}$~G (heavy curves) and $B=7\times 10^{13}$~G (light curves).  The photon energy is set to $E=0.5$~keV.}
\label{fig:stokes55}
\end{figure}

\begin{figure}
\centering
\includegraphics[scale=0.25]{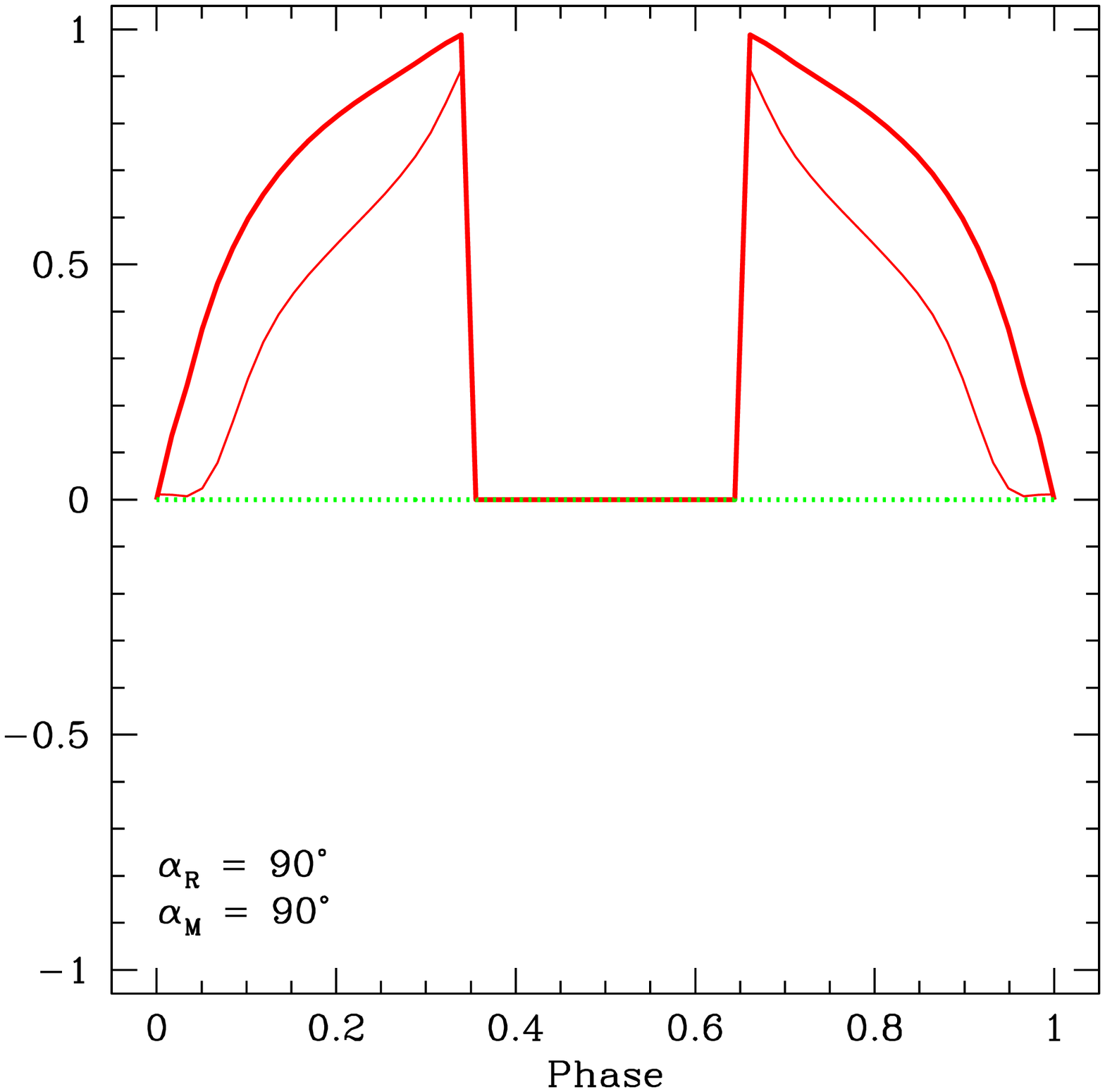}
\includegraphics[scale=0.25]{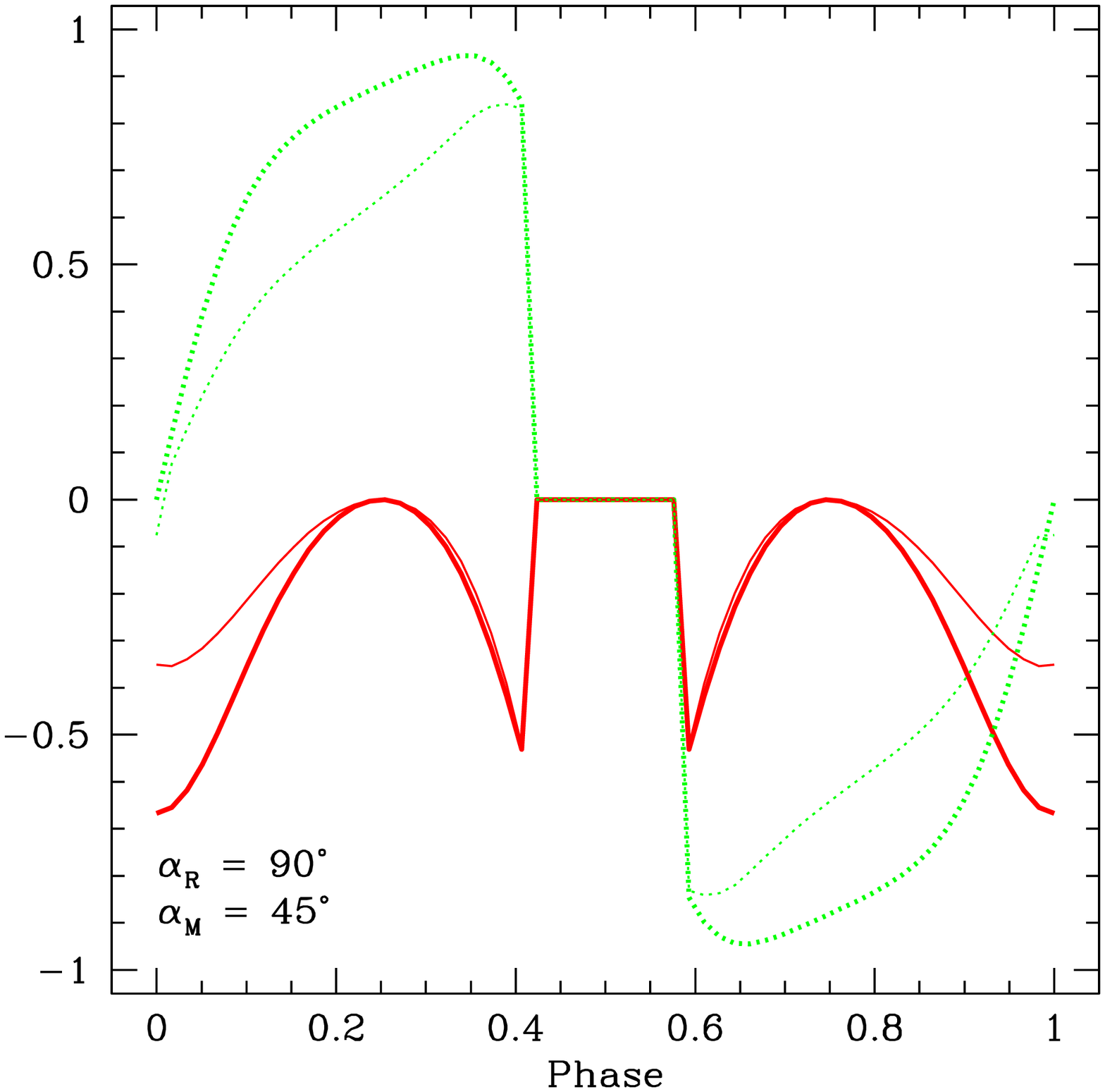}\\
\includegraphics[scale=0.25]{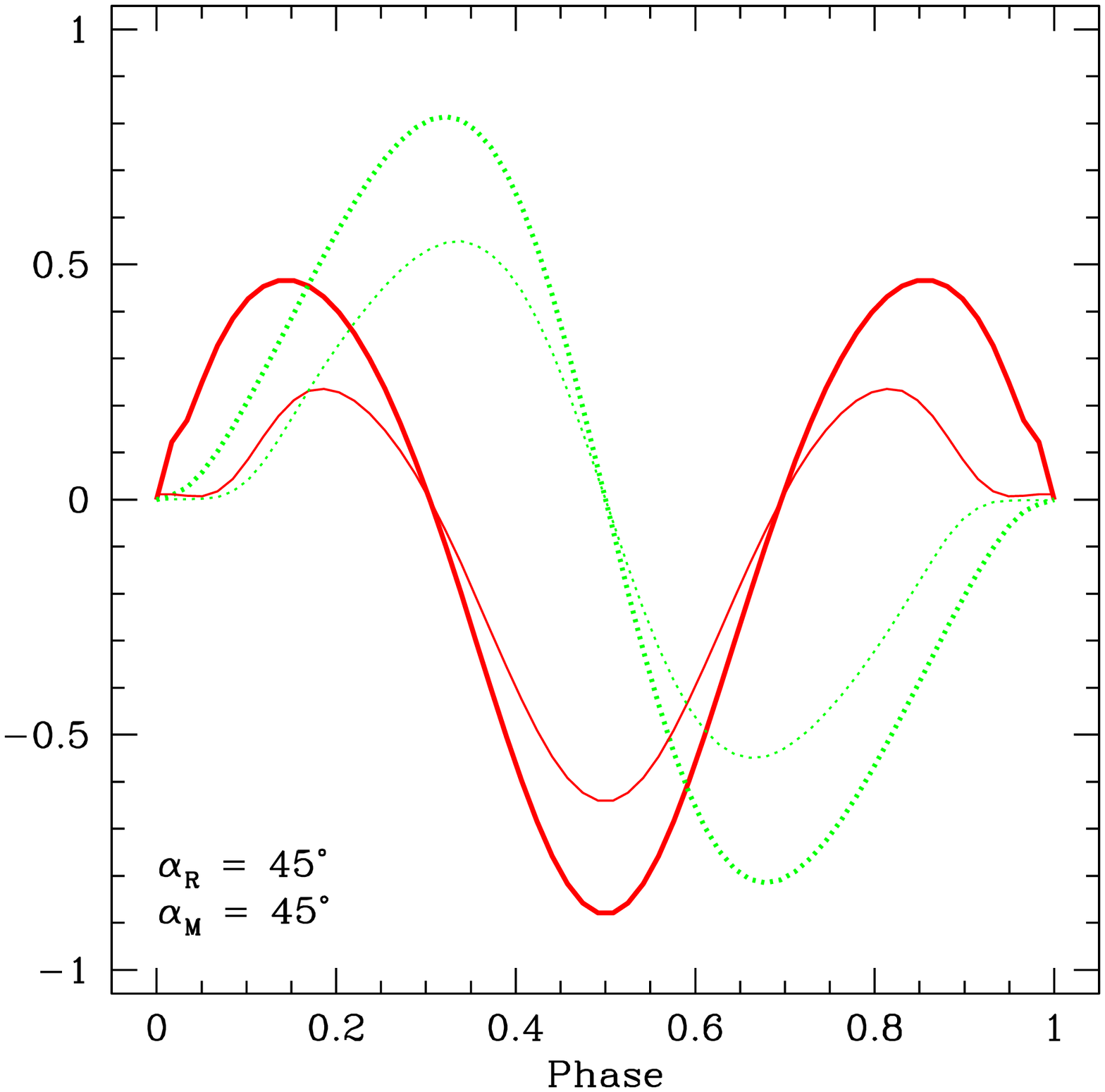}
\includegraphics[scale=0.25]{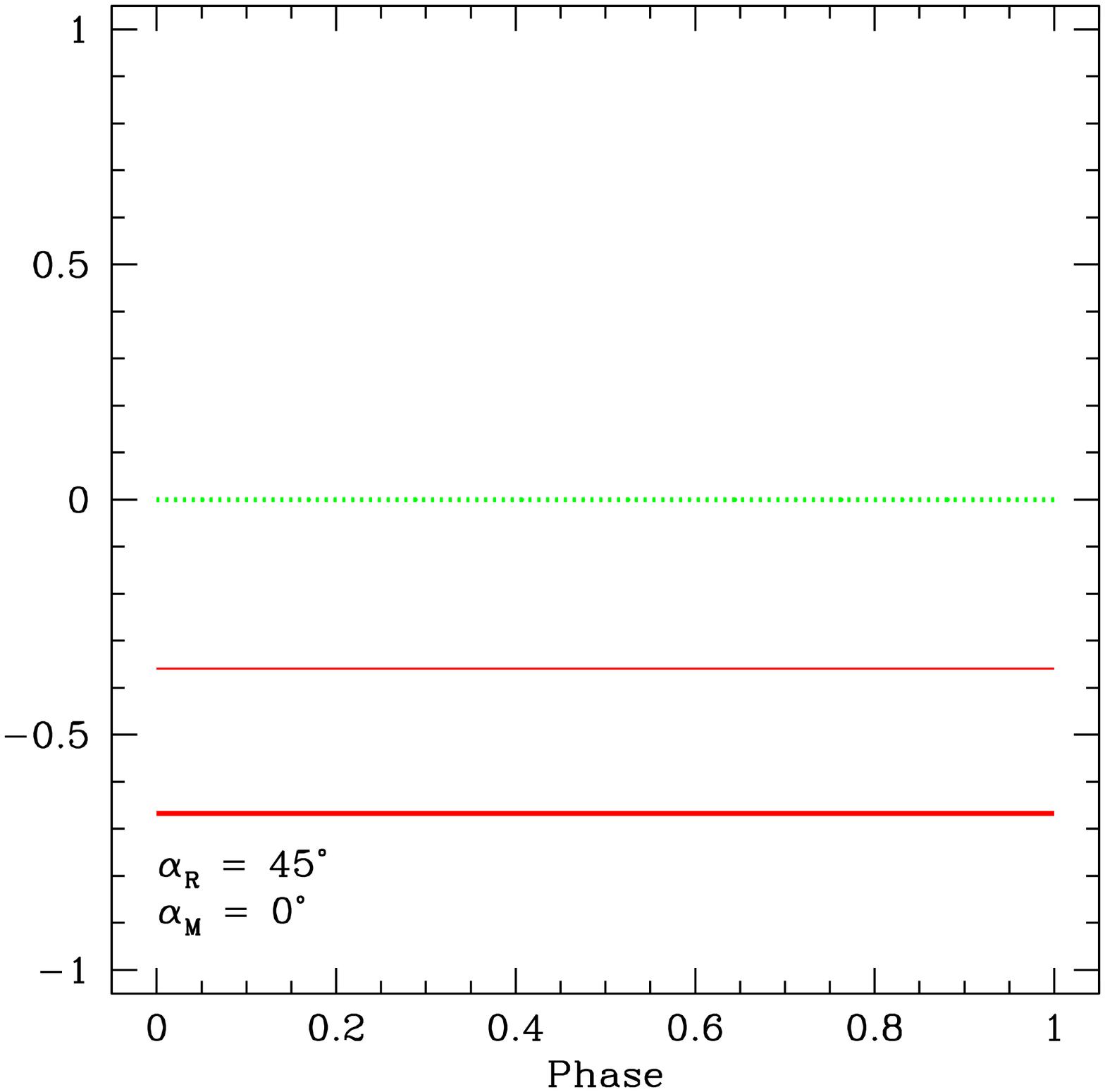}
\caption{Observed, normalized Stokes parameters $Q/I$ (solid curves), $U/I$ (dotted curves), as a function of phase, 
for an extended spot on a NS 
surface.  The spot has opening angle $\beta = 5^{\circ}$.  Results are shown for the same geometries as in 
figure~\ref{fig:pfracE5}, for magnetic field strengths 
$B=5\times 10^{14}$~G (heavy curves) and $B=7\times 10^{13}$~G (light curves).  The photon energy is set to $E=2$~keV.}
\label{fig:stokes205}
\end{figure}

\begin{figure}
\centering
\includegraphics[scale=0.25]{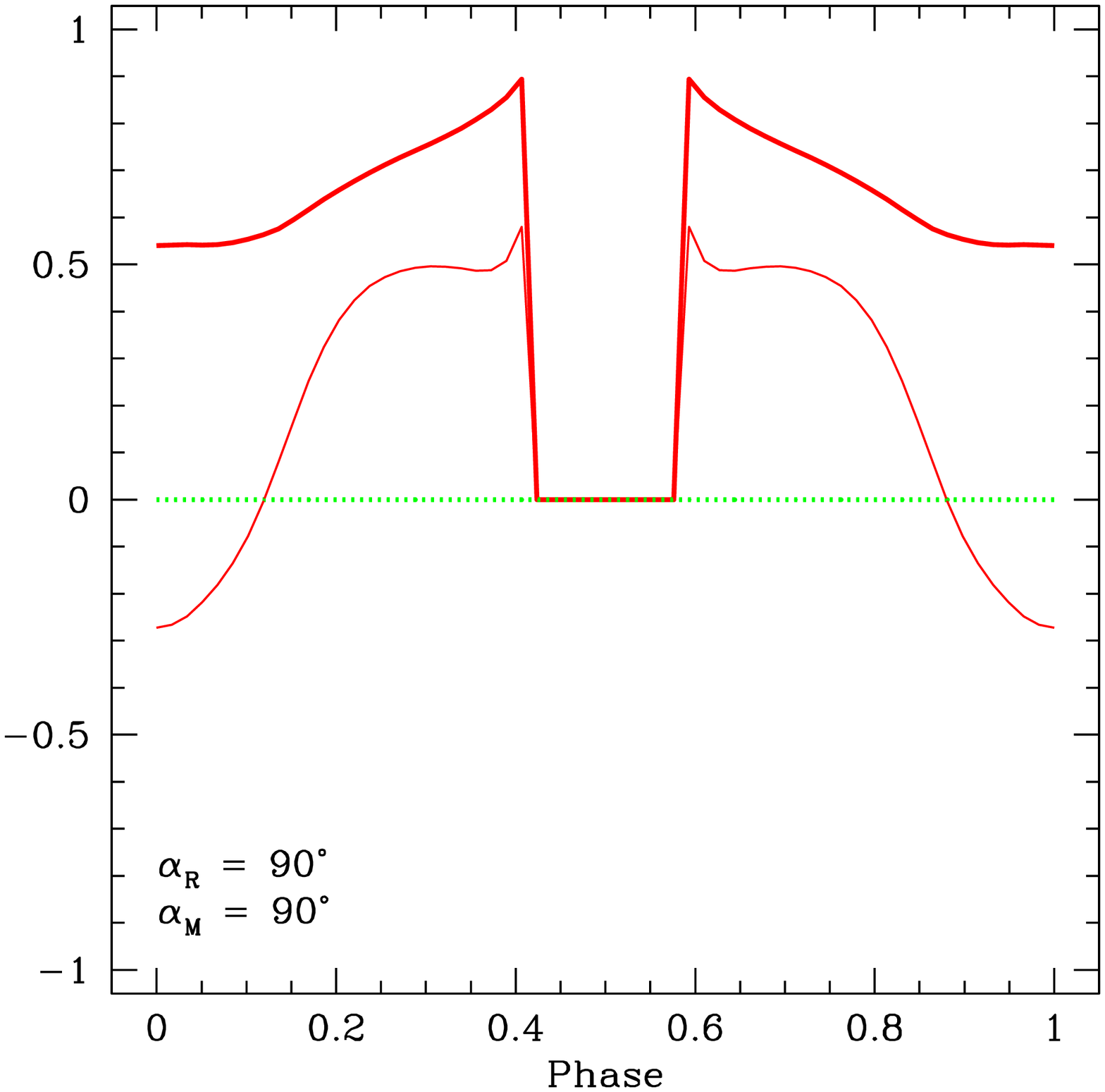}
\includegraphics[scale=0.25]{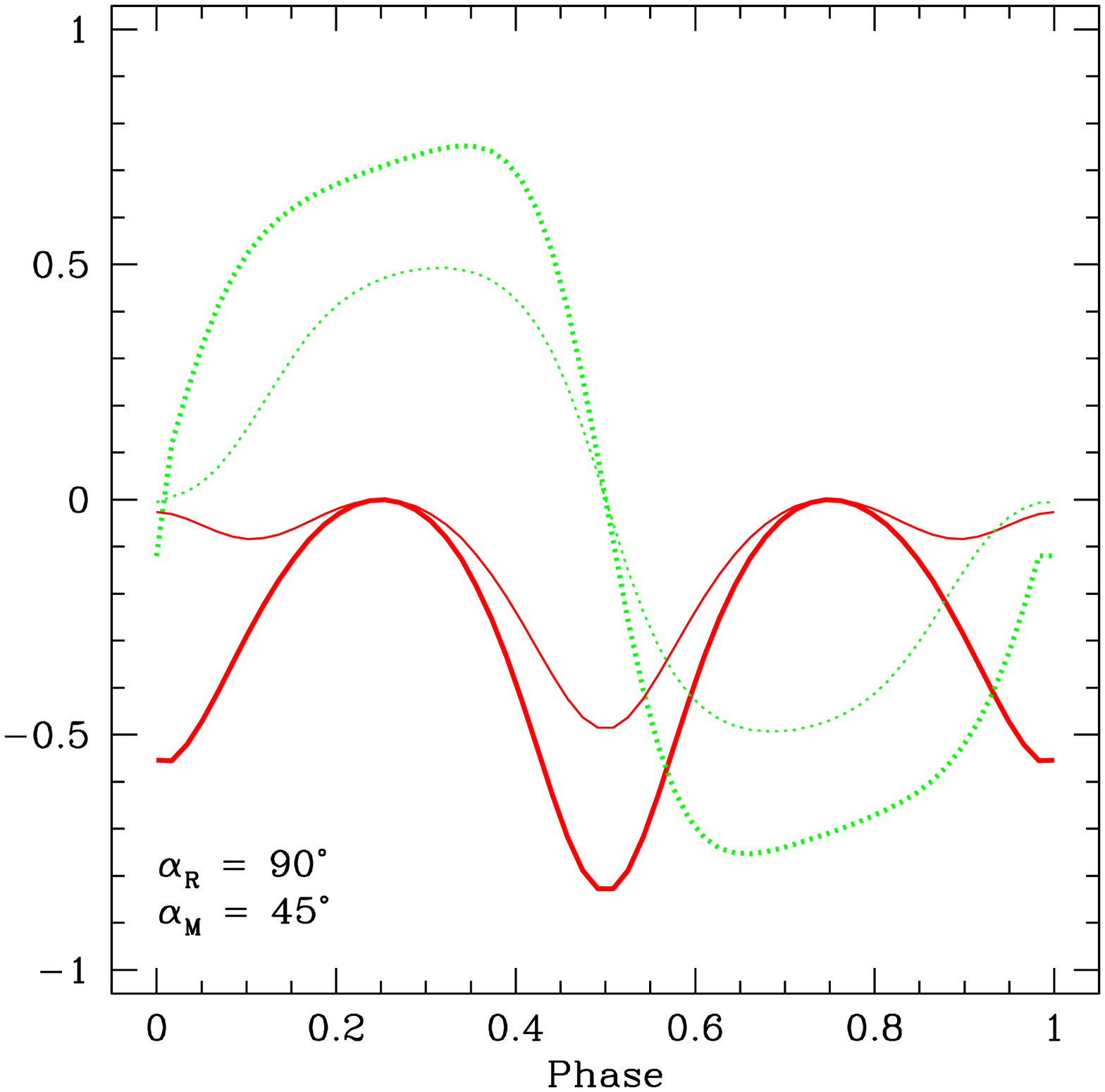}\\
\includegraphics[scale=0.25]{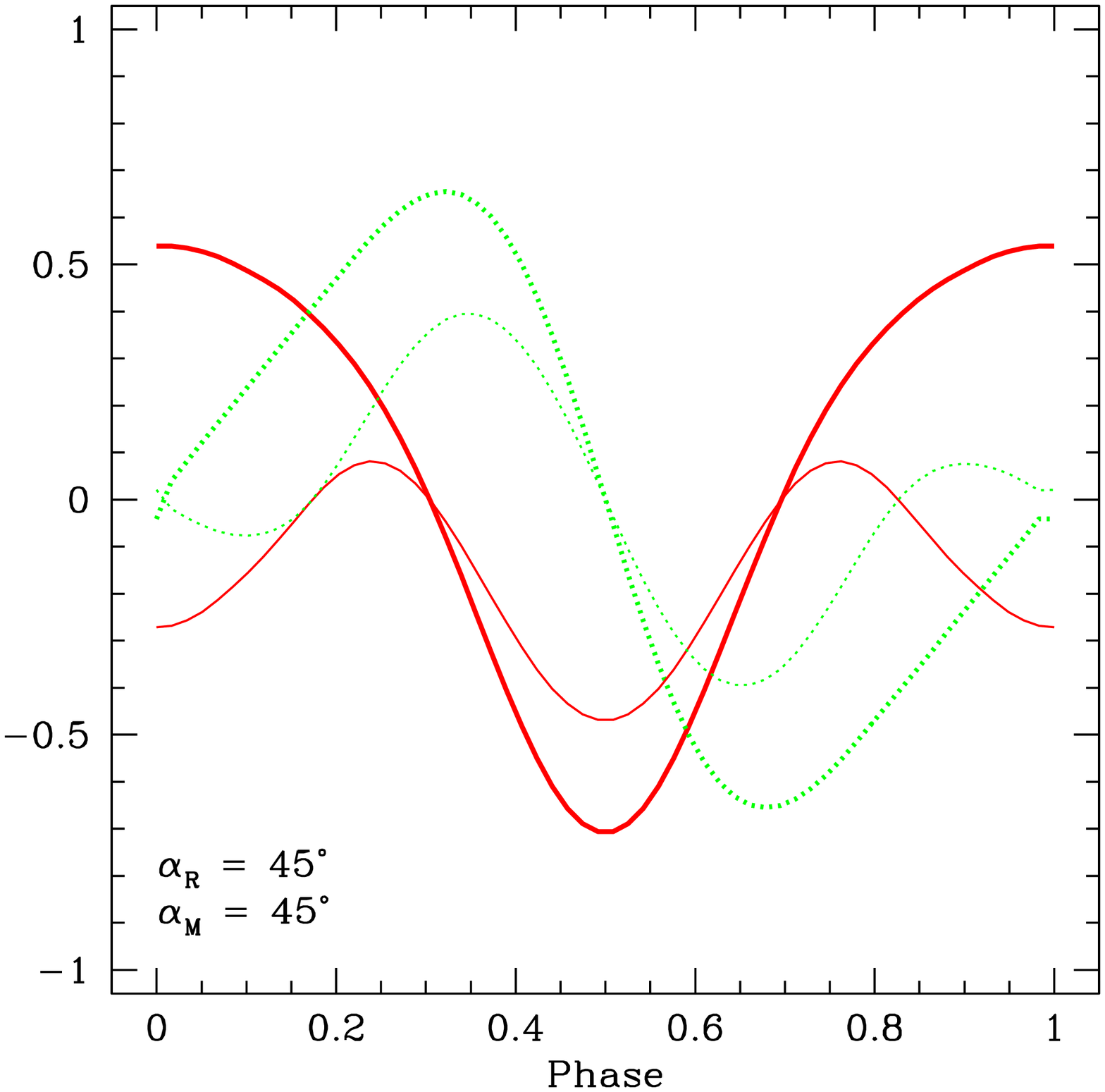}
\includegraphics[scale=0.25]{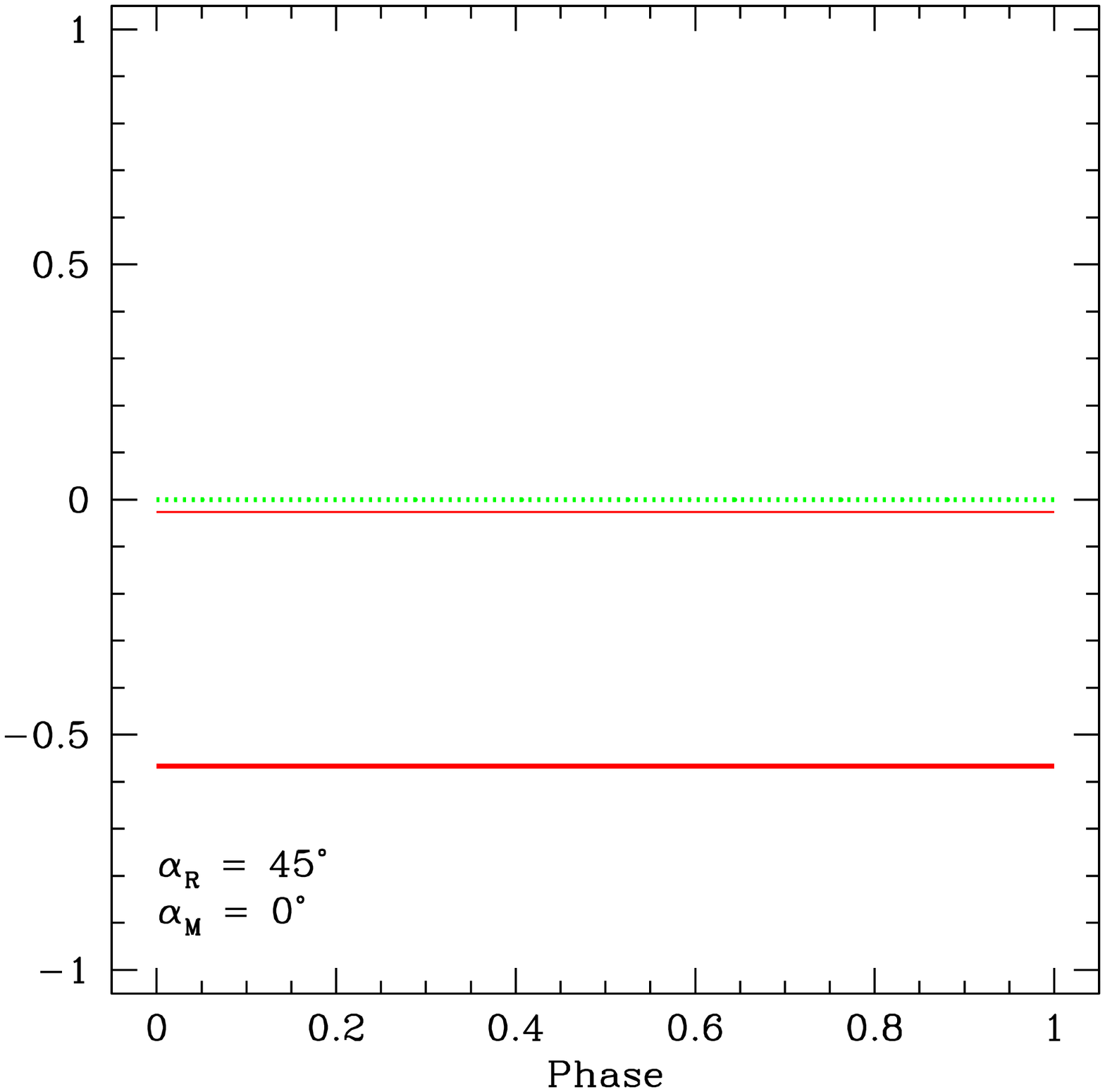}
\caption{Observed, normalized Stokes parameters $Q/I$ (solid curves), $U/I$ (dotted curves), as a function of phase, 
for an extended spot on a NS 
surface.  The spot has opening angle $\beta = 30^{\circ}$.  Results are shown for the same geometries as in 
figure~\ref{fig:pfracE5}, for magnetic field strengths 
$B=5\times 10^{14}$~G (heavy curves) and $B=7\times 10^{13}$~G (light curves).  The photon energy is set to $E=0.5$~keV.}
\label{fig:stokes530}
\end{figure}

\begin{figure}
\centering
\includegraphics[scale=0.25]{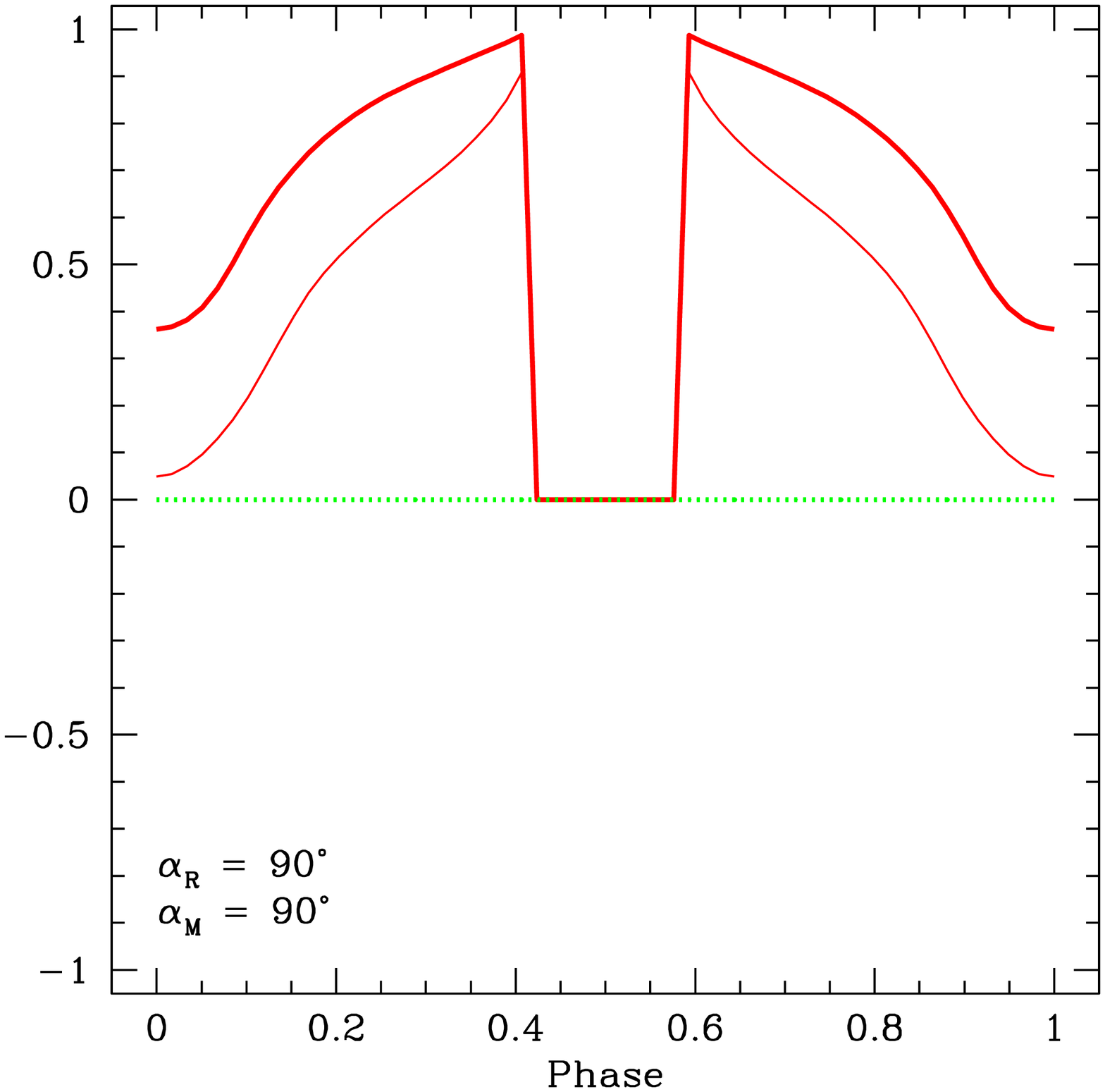}
\includegraphics[scale=0.25]{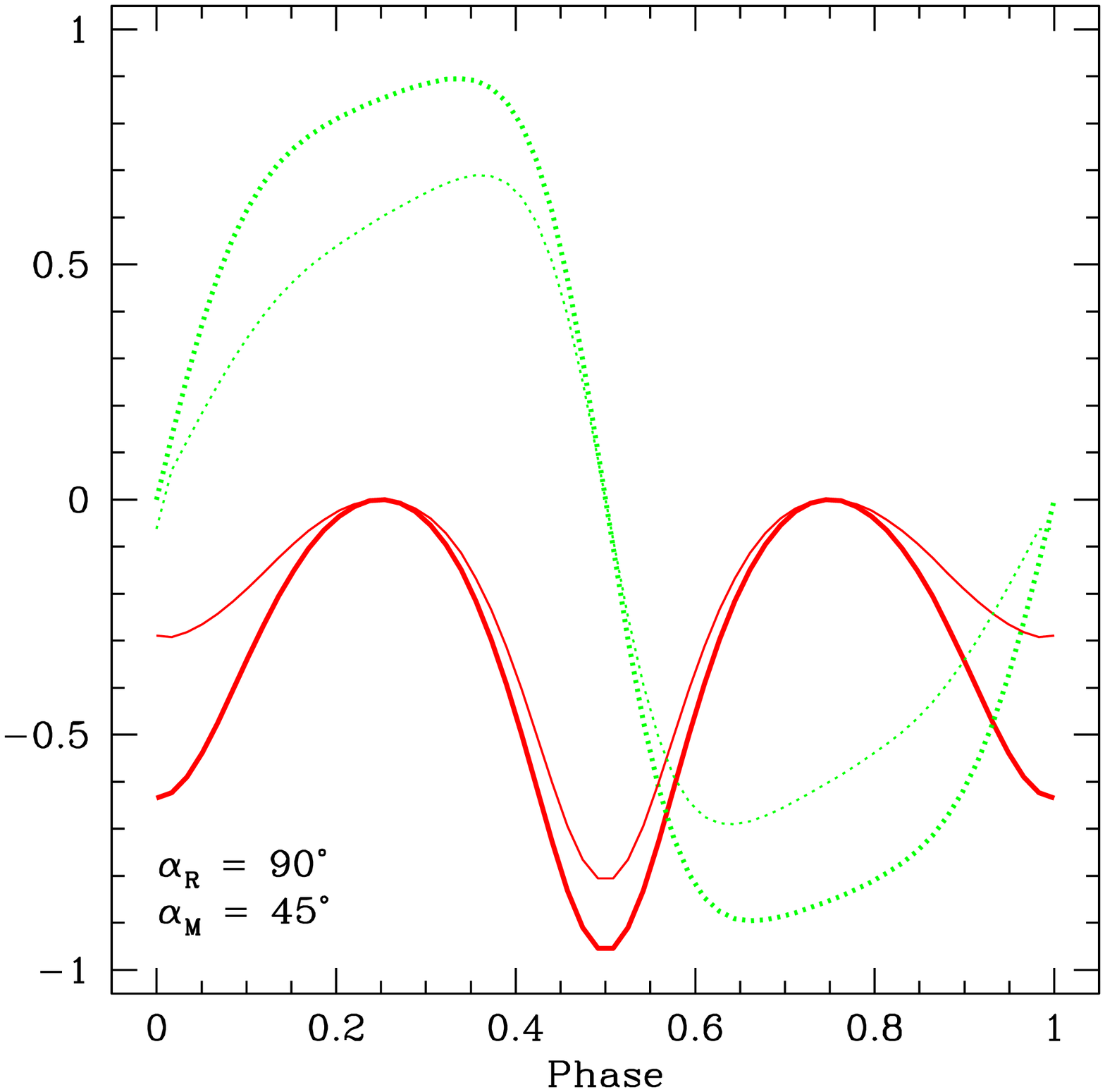}\\
\includegraphics[scale=0.25]{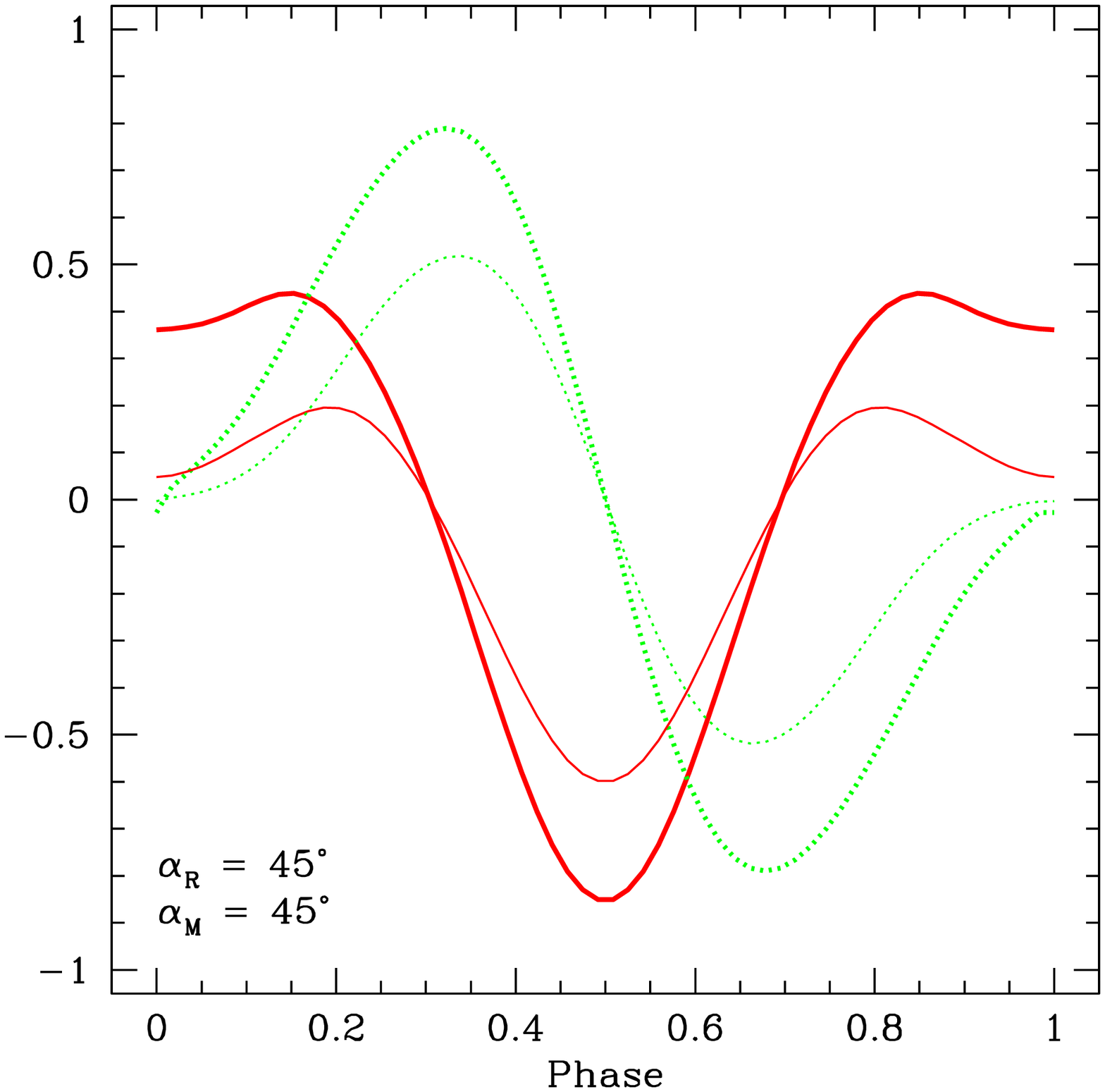}
\includegraphics[scale=0.25]{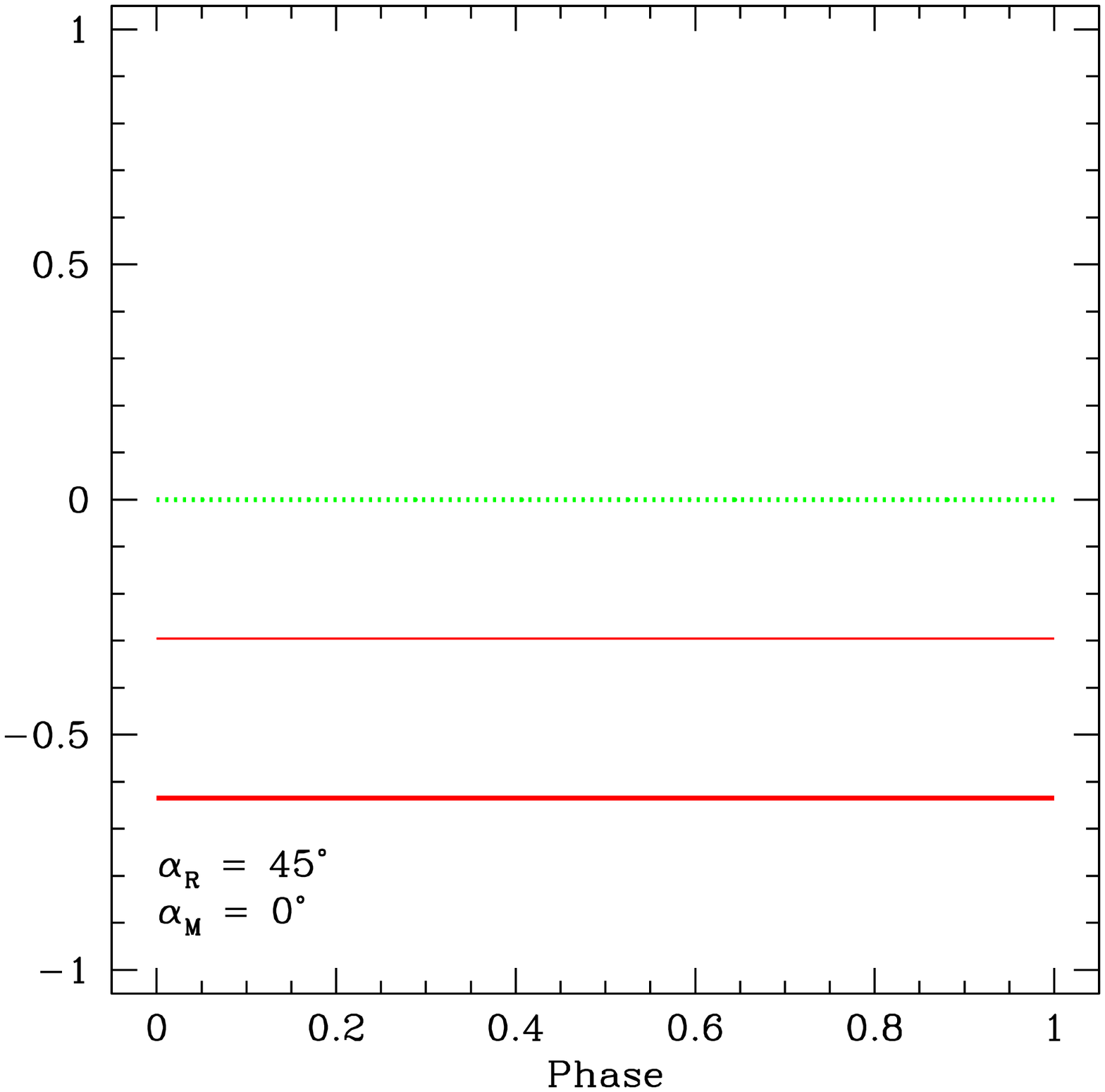}
\caption{Observed, normalized Stokes parameters $Q/I$ (solid curves), $U/I$ (dotted curves), as a function of phase, 
for an extended spot on a NS 
surface.  The spot has opening angle $\beta = 30^{\circ}$.  Results are shown for the same geometries as in 
figure~\ref{fig:pfracE5}, for magnetic field strengths 
$B=5\times 10^{14}$~G (heavy curves) and $B=7\times 10^{13}$~G (light curves).  The photon energy is set to $E=2$~keV.}
\label{fig:stokes2030}
\end{figure}

Figures~\ref{fig:stokes55}--\ref{fig:stokes2030} show the observed,
normalized photon Stokes parameters $Q_E/I_E$ (solid curves) and 
$U_E/I_E$ (dotted curves) at energies $E=0.5$~keV and $E=2$~keV as a
function of phase, for NS atmospheres with magnetic field strengths
$B=5\times 10^{14}$~G (heavy lines) and $B=7\times 10^{13}$~G (light
lines).  Each panel in the figures corresponds to one of the NS
geometries described above.  In
Figures~\ref{fig:stokes55} and \ref{fig:stokes205}, the opening angle of
the polar cap is set to $\beta=5^{\circ}$, while in
figures~\ref{fig:stokes530} and \ref{fig:stokes2030} it is set to
$\beta=30^{\circ}$.

In the case of the orthogonal rotator, 
$\sin(2\varphi_B) = 0$, and $\cos(2\varphi_B)=-1$, yielding $Q_E/I_E =
-\Pi^{\rm em}_E$, $U_E/I_E = 0$.  Thus, the projection of the magnetic field
into the $xy$ plane is along the $y$ axis.  In general, the projection
of the dipole magnetic field into the $xy$ plane far from the NS is approximately
equal to that of the magnetic dipole vector shifted by $\pi$ radians 
($\varphi\rightarrow \varphi_B + \pi$).
It is also important to note that the results are not
qualitatively sensitive to the polar cap size; the most significant difference
between Figures~\ref{fig:stokes55}, \ref{fig:stokes205} and
\ref{fig:stokes530}, \ref{fig:stokes2030} are the larger ranges of
phase over which the spot is visible.

\subsection{Phase-Averaged Polarimetry}

A unique signature of vacuum polarization effects on the evolution of photon modes was first reported by 
\citet[]{LaiHo03a}.  For NS with magnetic fields $B < 7\times 10^{13}$~G, the vacuum resonance density is less than the 
decoupling densities of X and O mode photons.  Thus, photons which have decoupled from the NS atmosphere 
undergo partial mode conversion at the vacuum resonance.  At low energies, $E\ll E_{\rm ad}$, and $\Pi_{\rm em}$ is 
unaffected.  However, at high energies, $E > E_{\rm ad}$, and the sign of $\Pi_{\rm em}$ reverses.  Thus, there is a 
rotation of the plane of linear polarization between low and high energy photons.

\begin{figure}
\centering
\includegraphics[scale=0.4]{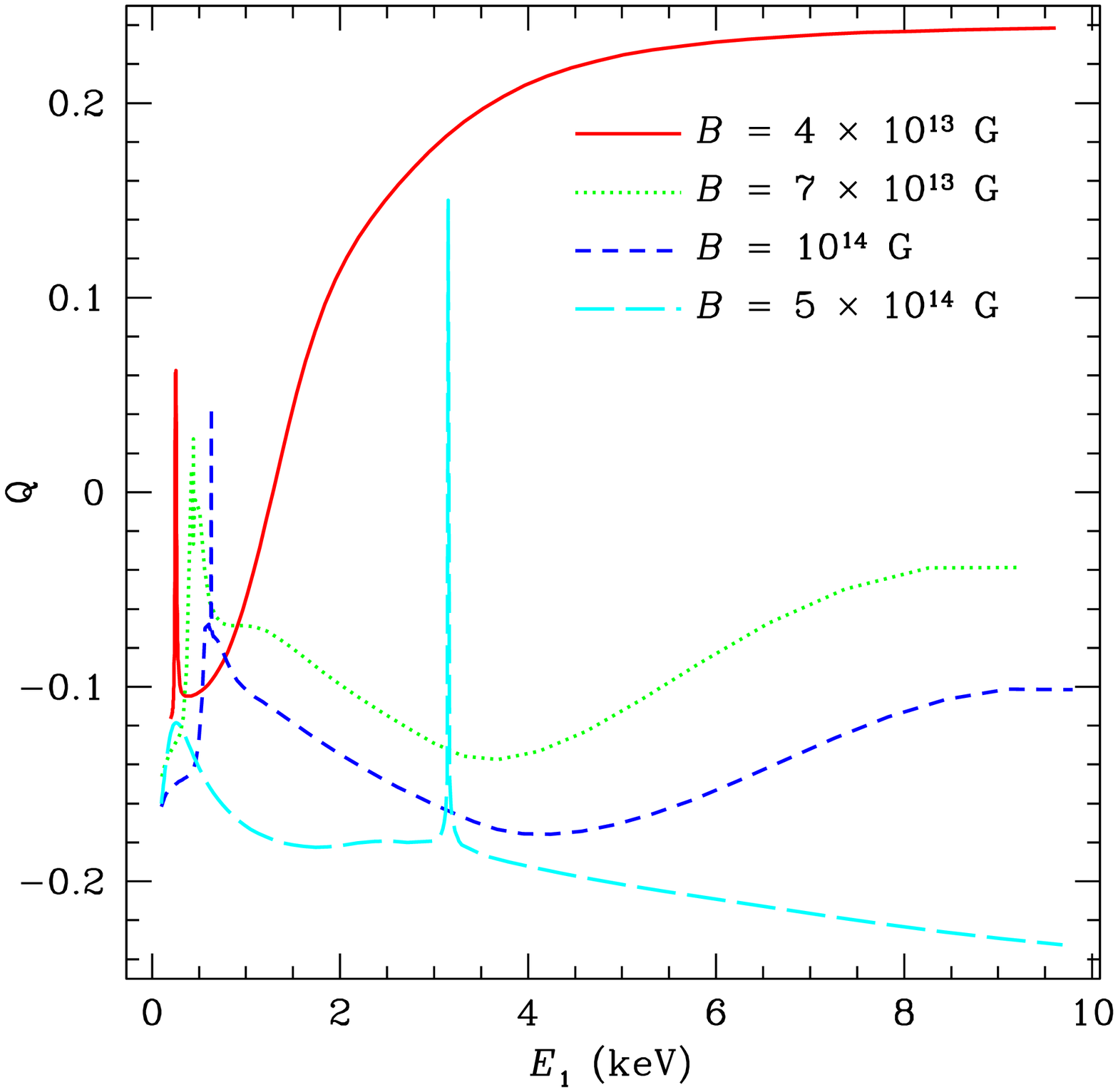}
\caption{Phase-averaged Stokes parameter $Q$ as a function of photon energy $E$ (keV).}
\label{fig:pave}
\end{figure}

In practice, this effect depends on the NS geometry (see \S\ref{subsect:EmPolFrac}).  For emission angles 
$\delta\approx 0$, $E_{\rm ad}\approx 0$ and no rotation between low and high photon energies is observed.  In geometries 
for which the typical emission angle is $\delta\approx \pi/4$, the effect can be pronounced.  Figure~\ref{fig:pave} shows 
the phase-averaged Stokes parameter $Q_{\rm ave}\equiv (2\pi)^{-1}\int_0^{2\pi}\,d\psi\, Q(\psi)$ as a function of photon 
energy for several magnetic fields with $B = 4 - 50\times 10^{13}$~G.  The geometry is case G2 with opening angle 
$\beta = 5^{\circ}$.  Note that $U_{\rm ave}$ vanishes by symmetry.
For $B=4\times 10^{13}$~G (solid curve), $Q_{\rm ave}$ changes sign at $E\approx 1.5$~keV, the energy at which 
resonant mode conversion becomes effective.  For $B = 7\times 10^{13}$~G, the vacuum resonance density is nearly equal to 
the O mode decoupling density, depending on the NS rotational phase.  For most values of $\psi$, $\rho = \rho_V$ is within 
the X and O mode photospheres, and no rotation of the plane of polarization occurs.  Thus, $Q_{\rm ave}$ retains the same 
sign at low and high energies \citep[c.f., figure~20 of][for a similar result in a different geometry]{vanAdelsbergLai06a}.

\subsection{Polarization, Atmosphere Composition, and NS Equation of State}

Magnetar atmospheres are typically assumed to consist of mostly hydrogen. However, if hydrogen is depleted by 
thermonuclear burning in the photosphere, they may consist of helium \citep[][]{Changetal04a}.
For fully ionized helium composition at $B=5\times 10^{14}$~G, and $T_{\rm eff} = 5\times 10^6$~K, we find that 
the observed Stokes parameters differ from those in the hydrogen case by 
$\la 10\%$ for $0.1 < E < 10$ keV.  We therefore do not show results for this case.  The reason for the similarity is that 
the O mode decoupling depth in ionized helium is $\sim 60\%$ of that in hydrogen, while the vacuum resonance density 
(identified with the X mode decoupling depth at large magnetic field strengths) increases by a factor of 
two.  The temperature profile evolves much more slowly than the density; in an interval in which the density increases by 
several orders of magnitude, the temperature only changes by a factor of a few.  Thus, at most energies and 
propagation angles, the relative decoupling 
temperatures of X and O photons are similar to those in the hydrogen case, and we expect to see the same qualitative 
features, with a slightly larger magnitude of the helium polarization fraction.  
Nevertheless, at lower fields, the reduction of the O mode decoupling depth and increase in the 
resonance density supresses the rotation of the plane of polarization between low and high energy photons at 
$B \approx 4\times 10^{13}$~G.  For helium, lower magnetic fields, which are unlikely to occur in the magnetar regime,
are necessary to see this effect.

We also explored the dependence of the emitted polarization
fraction on the equation of state of the star. In particular, for the
case of $B=5\times 10^{14}$~G and $\beta=5^\circ$, we compared the
polarization fraction for three values of the NS radius, $R=9,12,15$~km 
with the NS mass fixed at $M=1.4 M_\odot$.  For the four geometries defined
above, we found that the range of $\psi$ for which the spot is visible increases with
decreasing radius.  This result is expected, due to light deflection effects around
relativistic stars.  However, the difference between the emitted
polarization fraction for the two extreme radii is
$|\Pi_E^{\rm em}(R=15\;{\rm km})-\Pi_E^{\rm
em}(R=9\;{\rm km})|/|\Pi_E^{\rm em}(R=15\;{\rm km})|\ll 1$, for most values of $\psi$.  
Since the phased flux modulation 
of the X and O mode photons is equally affected by changes in the NS radius,
the variation in the linear polarization fraction with
$R$ is associated with two effects that vary with NS compactness: 
(1) the difference between the energy at
the star surface and the observed energy, and (2) the
range of emission angles $\delta$ for which photons reach the observer
at a given rotational phase $\psi$.  The ratio of emitted energies
corresponding to the same observed energy for stars with $R=9$~km and $R=15$~km 
is $E_9/E_{15}\approx 1.15$.  Similarly, the difference in the range of $\delta$ for photons 
which reach the observer, at a given $\theta$, is smaller than the spot size for small
$\theta$, and, for most geometries, is generally smaller than the range
over which substantial polarization variations occur.
Therefore, the polarization signal has no significant, direct dependence on the
EOS of the NS.

\section{Discussion}
\label{sect:Discussion}

Current studies of NSs rely on spectroscopy to constrain the stars' physical parameters.  
Fits to the phase-averaged X-ray spectrum yield the effective temperature of the
star and the size of the emitting region, but are 
insensitive to the viewing geometry.  Timing analysis complements the
information obtained from spectral fits by constraining
the permitted values of $\alpha_R$ and 
$\alpha_M$.  However, since the amplitude of the modulation of the flux with
phase is largely dependent on the mass-to-radius ratio of the star,
there is a degeneracy between the inferred viewing geometry and compactness ratio 
\citep[e.g.,][]{PernaGotthelf08a}.

As shown above, our calculations of the magnetar polarization signal are not sensitive to $M/R$, and can therefore be 
used to break this degeneracy.  Moreover, the sensitivity of the Stokes parameters to the magnetic field strength can be 
used to constrain the magnetar magnetic field.  In particular, the phase-averaged behavior of the Stokes parameters at 
low and high photon energies in the soft X-ray band can put lower or upper limits on $B$ (in the absence or 
presence of rotation of the plane of linear polarization, respectively).  

Measurements of the magnetic field of a magnetar from the shape of the
continuum X-ray spectrum alone are extremely difficult to perform.
Magnetic atmosphere models are characterized by hard tails with
respect to blackbody emission \citep[this discrepancy is larger for
magnetic fields $B<B_Q$; see][]{HoLai03a,vanAdelsbergLai06a}.
Spectral fits are generally unable to uniquely discriminate among
different B-field strengths based on the spectral shape alone,
especially since hard tails are also predicted from scattering in the
atmosphere \citep[][]{Reaetal08a}.  While no features have been
detected in quiescent magnetar thermal spectra, there is one confirmed
detection of an absorption-like feature in an outburst of the
Anomalous X-ray Pulsar 1XRSJ170849-400910 \citep[][]{Reaetal03a}.  The
phase-dependent feature occurs at 8.1~keV in the magnetar spectrum,
consistent with resonant cyclotron scattering by electrons at $9\times
10^{11}$~G or ions at $1.6\times 10^{15}$~G.  The value of the magnetic
field inferred from dipole spin-down is $B\sim 5\times 10^{14}$~G,
however, much higher surface fields are possible if the magnetic field
is dominated there by higher-order multipoles.  Nevertheless, an unambiguous
identification of the line feature, and hence, magnetic field
strength, is not possible.  Future measurements of Stokes parameters
in the soft X-ray band may help break this degeneracy.
If the rotation of the plane of linear polarization is not observed,
this is consistent with a the magnetar in the strong magnetic field
regime, and the identification of the line feature with ion cyclotron
resonance becomes more certain.

\section*{Acknowledgements}

We thank Tim Kallman for motivating this work, and Dong Lai
for very useful comments on the manuscript.

\bibliography{polarization}

\end{document}